\documentclass[12pt]{article}
\pdfoutput=1

\usepackage[a4paper,text={16.8cm,22.4cm}]{geometry}
\usepackage{amsmath,amsfonts,braket,slashed,amssymb,tikz,bm,psfrag,graphicx,color,dsfont,tikz}
\usepackage{multicol}
\usepackage{ctable}

\usepackage{cite}

\usepackage[
	colorlinks=true,
	urlcolor=blue,
	linkcolor=red,
	citecolor=blue
]{hyperref}

\begin{document}

\vspace{1.0cm}

\begin{center}
\Large\bf\boldmath
Universal Properties of Pseudoscalar Mediators
\end{center}

\vspace{0.2cm}
\begin{center}
Martin Bauer$^a$, Martin Klassen$^a$ and Valentin Tenorth$^{a, b}$ \\
\vspace{0.7cm} 
{\sl ${}^a$Institut f\"ur Theoretische Physik, Universit\"at Heidelberg
Philosophenweg 16, 69120 Heidelberg, Germany}\\
{\sl ${}^b$Max-Planck-Institut f\"ur Kernphysik (MPIK),
Saupfercheckweg 1, 69117 Heidelberg,}\\[.4cm]
\textbf{Abstract}\\[3mm] %%
\parbox{0.9\textwidth}{We discuss universal signals of consistent models of pseudoscalar mediators for collider searches for Dark Matter. Keeping only the degrees of freedom that can not be decoupled due to consistency conditions, we present a universality class of simplified models with pseudoscalar mediators and renormalizable couplings to Standard Model fields. We compute stability and perturbativity constraints, constraints from electroweak precision measurements, collider searches for new heavy particles as well as constraints from relic density measurements and indirect detection experiments searching for signals of Dark Matter annihilation into photons. We find that the mono-$Z$ final state is the strongest, universal signal of this class of models, with additional signatures present in the different ultraviolet completions that can be used to distinguish between them.  
}
\end{center}

\vspace{0.cm}

\section{Introduction}\label{sec:intro}
Extensions of the Standard Model (SM) with a single mediator and a Dark Matter candidate provide an important tool for Dark Matter searches at colliders, since they capture the kinematics of on-shell propagators and allow to derive results which generalize to a large class of more complete theories of the dark sector \cite{Fox:2012ee, Harris:2014hga, Baek:2015lna, DeSimone:2016fbz, Bauer:2016pug}. Simplified models typically feature neutral mediators with a single dark matter particle, see \cite{Abdallah:2014hon,Abdallah:2015ter, Abercrombie:2015wmb, Albert:2017onk} and references therein. Couplings between the mediator and SM fermions are constrained by flavour observables \cite{Dolan:2014ska}, such that spin-1 mediators are expected to have universal couplings to all flavours of a given charge and spin-0 mediators couplings are expected to have Yukawa-like structures \cite{Kahlhoefer:2015bea}. This leads to strong constraints on spin-1 mediators from di-lepton and di-jet searches, resulting in better limits than mono-X searches over a large range of the perturbatively allowed parameter space \cite{Fairbairn:2016iuf, Okada:2016gsh,  summaryplots}, while couplings of spin-0 mediators to leptons and light quarks are suppressed by their masses. Direct detection experiments further constrain spin-0 mediators with scalar couplings to SM fermions, whereas spin-0 mediators with pseudoscalar couplings lead to velocity suppressed couplings between nuclei and dark matter \cite{Zheng:2010js}. This makes collider searches particularly powerful in constraining pseudoscalar mediators.  \\
Renormalizable couplings of gauge singlet pseudoscalars to SM fermions break the electroweak gauge symmetry. This breaking manifests itself in unitarity violating amplitudes, signaling the breakdown of these models in the absence of additional states. By considering such simplified models, it is therefore implicitly assumed that the couplings result from a gauge-invariant scalar sector upon integrating out some of the additional states. However, it has been shown in explicit constructions of gauge invariant models that these additional states can not be arbitrarily heavy without leading to tension with measurements of electroweak precision observables \cite{Goncalves:2016iyg, Bauer:2017ota}. The presence of extra states in the vicinity of the pseudoscalar mediator imply interesting additional signatures of pseudoscalar mediator models. In particular the hierarchy of constraints from mono-X searches from initial-state radiation
\begin{align*}
\text{mono-jet} > \text{mono-photon} > \text{mono-$Z/W^\pm$} > \text{mono-Higgs}
\end{align*}
can be broken up by resonantly enhanced mono-$Z$, mono-$W^\pm$ and mono-Higgs final states \cite{ Bauer:2017ota}. Consistent, gauge invariant models of pseudoscalar mediators are however not unique. Whether some of these final states are enhanced or additional signatures arise therefore depends on the specific way in which gauge invariance is restored in the full theory. In this paper, we work out signals common to a large class of gauge invariant models, by exploring a consistent, gauge invariant effective field theory (EFT) in which only those additional mediators that cannot be decoupled are kept as explicit degrees of freedom. We establish the parameter space in which the dark matter candidate can explain the observed relic density in this class of models, compute constraints from direct detection experiments taking the additional mediators into account and finally provide a strategy for collider searches for this well-motivated region of parameter space.\\
The reminder of this paper is structured as follows. In Section \ref{sec:models}, we discuss the different consistent, gauge-invariant simplified models for pseudoscalar mediators, and present the EFT on which the analysis is based. In Section \ref{sec:constraints1} and Section \ref{sec:constraints2}, we present constraints on the parameter space of this EFT from measurements of Higgs couplings, electroweak precision parameters, the observed relic density and indirect detection experiments, before we turn to collider search strategies in Section \ref{sec:collider} and conclude in Section \ref{sec:conclusion}.  

\section{Consistent Models for Pseudoscalar Mediators}\label{sec:models}
Gauge invariant models of pseudoscalars imply new particles charged under the SM gauge symmetries.\footnote{A new spin-0 singlet with pseudoscalar couplings to Dark Matter can mix with the SM Higgs, which will lead to spin-0 mass eigenstates which are not CP eigenstates. Such a scenario does not require new states beyond the mediator and dark matter, but is strongly constrained by Higgs coupling measurements \cite{Baek:2017vzd}.} 
If the pseudoscalar mediator is a SM singlet, the couplings between $SU(2)_L$ singlet quarks $q_i=u_i, d_i$ and leptons $\ell_i$, as well as $SU(2)_L$ doublet quarks $Q_i=(u_i,d_i)$ and leptons $L_i=(\nu_i,\ell_i)$, for all three generations $i=1,2,3$ to a Dirac-fermion $\chi=\chi_L+\chi_R$ dark matter candidate can be described by the effective couplings 
\begin{align}\label{eq:SEFT}
\mathcal{L}= \sum_{i,j=1}^3 y_{ij}^q\frac{a}{\Lambda} \bar Q_i\,\gamma_5   \,q_j\, H+ \sum_{i,j=1}^3 y_{ij}^\ell\frac{a}{\Lambda} \bar L_i\,\gamma_5   \,\ell_j\,H+ c_s\,a\,\bar\chi\gamma_5\chi \,+h.c..
\end{align}
Here, $H$ denotes the $SU(2)_L$ doublet Higgs boson and the scale $\Lambda$ is typically associated with the mass of additional color-charged fermions or additional inert scalar doublets \cite{Calibbi:2012at}. Searches for these additional particles put strong constraints on the mass scale $\Lambda$, which leads to a suppressed production cross section of the pseudoscalar $a$ at colliders. Further, some form of minimal flavour violation needs to be generated by the ultraviolet (UV) theory to evade strong bounds from flavour violating neutral currents. If the pseudoscalar $a$ is the component of an $SU(2)_L$ doublet instead, the corresponding EFT reads (assuming Yukawa couplings of a two-Higgs doublet model of type II here)
\begin{align}\label{eq:2HDMEFT}
\mathcal{L}= \sum_{i,j=1}^3y^u_{ij} \bar Q_i\,H_1  \,u_j\, +\sum_{i,j=1}^3y^d_{ij} \bar Q_i\,H_2  \,d_j\, +  \sum_{i,j=1}^3y^\ell_{ij} \bar L_i\,H_2  \,\ell_j+\, c_\chi\frac{H_1^\dagger H_2}{\Lambda}\,\bar\chi\chi \,+c_5\frac{H_1^\dagger H_2}{\Lambda}\,\bar\chi \gamma_5\chi +h.c.\,,
\end{align}
where the scale $\Lambda$ is associated with new states not necessarily charged under color or $SU(2)_L$, and is therefore less constrained. In contrast to \eqref{eq:SEFT}, the coupling to SM fermions are renormalizable, while the coupling of the mediator to Dark Matter is suppressed. The production cross section $\sigma(pp\to a)$ can therefore be large, because the coupling to top-quarks can be of order one, while the branching ratio of the pseudoscalar $\text{Br}(a\rightarrow \chi\chi)$ can still dominate, due to the $m_f/v$ suppression of its couplings to SM fermions.   
This makes this class of models very interesting for collider searches.
In \eqref{eq:2HDMEFT} natural flavour conservation is assumed and the couplings to dark matter $c_\chi, c_5$ are in general complex couplings. Operators of the type $H_i^\dagger H_i \bar\chi \chi $, $i=1,2$, which do not induce pseudoscalar couplings are not explicitly included. This omission can be justified if one considers a new softly broken symmetry under which the SM singlets $H_1^\dagger H_2$, $\bar \chi \chi$ and $\bar \chi \gamma_5 \chi$ are charged \cite{Bauer:2015fxa, Bauer:2015kzy}. We will however not constrain the discussion to this case. \\
The effective Dark Matter couplings in \eqref{eq:2HDMEFT} can be obtained in different well-motivated UV completions, by considering the additional states heavy with respect to the SM particles, the scalar and pseudoscalar components of the Higgs multiplets and the Dark Matter candidate. One example is a UV completion in which a SM singlet pseudoscalar mediator mixes with the combination $H_1^\dagger H_2$ \cite{Goncalves:2016iyg, Bauer:2017ota},
\begin{align}\label{eq:UVcomp1}
\mathcal{L}= \sum_{i,j=1}^3y^u_{ij} \bar Q_i\,H_1  \,u_j\, +\sum_{i,j=1}^3y^d_{ij} \bar Q_i\,H_2  \,d_j\, +\sum_{i,j=1}^3y^\ell_{ij} \bar L_i\,H_2  \,\ell_j \,+ \kappa\,a\,H_1^\dagger H_2+c_a\,a\,\bar\chi \gamma_5\chi \,+h.c.\,.
\end{align}
Another UV completion arises from more complicated dark sectors, with additional electroweak doublet fermions $\psi=(\chi^+,\chi^0)^T$,
\begin{align}\label{eq:UVcomp1}
\mathcal{L}= \sum_{i,j=1}^3y^u_{ij} \bar Q_i\,H_1  \,u_j\, +\sum_{i,j=1}^3y^d_{ij} \bar Q_i\,H_2  \,d_j\, +\sum_{i,j=1}^3y^\ell_{ij} \bar L_i\,H_2  \,\ell_j \,+c_1\,\bar\psi H_1\chi +c_2\,\bar\psi \tilde H_2\chi+h.c.\,,
\end{align}
as in extended doublet-singlet dark matter models \cite{Berlin:2015wwa}, encompassing the Bino-Higgsino limit of the minimal supersymmetric Standard Model. While these UV completions predict very different, model-specific signatures that allow to differentiate between them, the focus of this work is on universal signals that arise in \emph{all} pseudoscalar mediator models which lead to the EFT \eqref{eq:2HDMEFT}.\\

\section{Higgs Couplings, Stability Constraints and Collider Searches for Heavy Resonances }\label{sec:constraints1}

The parameter space of the effective theory described in the previous section is constrained by observables, independent of the couplings to dark matter, such as flavour and electroweak precision observables, Higgs coupling measurements, and searches for the additional scalars decaying in SM final states. In the following, we discuss the corresponding constraints on the parameter space.

\subsection{Higgs Couplings }\label{sec:Higgscouplings}

The effective theory described by the Lagrangian \eqref{eq:2HDMEFT} in combination with the scalar potential (again assuming a global symmetry that is softly broken by the real parameter $\mu_3$)
\begin{align} \label{eq:VH}
V_{H} & = \mu_1 H_1^\dagger H_1 + \mu_2 H_2^\dagger H_2 + \left ( \mu_3  H_1^\dagger H_2 + h.c. \right ) + \lambda_1  \hspace{0.25mm} \big ( H_1^\dagger H_1  \big )^2  + \lambda_2  \hspace{0.25mm} \big ( H_2^\dagger H_2 \big  )^2 \notag\\
& \phantom{xx} +  \lambda_3 \hspace{0.25mm} \big ( H_1^\dagger H_1  \big ) \big ( H_2^\dagger H_2  \big ) + \lambda_4  \hspace{0.25mm} \big ( H_1^\dagger H_2  \big ) \big ( H_2^\dagger H_1  \big )  \,,
\end{align}
introduces ten new parameters, the mass of the dark matter candidate $m_\chi$ and its couplings $c_\chi, c_5$, as well as three dimensionful and four dimensionless parameters $\mu_1,\mu_2,\mu_3, \lambda_1, \lambda_2, \lambda_3, \lambda_4$. The latter can be traded for the expectation values of the two Higgs doublets $v_1$ and $v_2$ or $v =\sqrt{v_1^2+v_2^2}=246$ GeV and $t_\beta=\tan\beta=v_1/v_2$, the masses of the physical scalars $M_h=125$ GeV, $M_H, M_{H^\pm}$, and of the pseudoscalar $M_A$, and the scalar mixing angle $c_{\beta-\alpha}=\cos(\beta-\alpha)$ (or $s_{\beta-\alpha}=\sin(\beta-\alpha)$). We rewrite  \eqref{eq:2HDMEFT} such that the couplings of the neutral scalars read
\begin{align}
\mathcal{L}= \sum_{f=u, d, \ell}\sum_{j=1}^3 g_{h f}\,\bar f_j f_j \,h + g_{H f}\,\bar f_j f_j \,H +ig_{A f}\,\bar f_j \gamma_5 f_j \,A+ \!\!\!\!\sum_{\varphi=h,H,A}g_{\varphi \chi} \bar \chi\chi \,\varphi + ig_{\varphi 5} \bar \chi \,\gamma_5\chi \,\varphi\,.
\end{align}
Couplings between SM fermions $f$ and the neutral scalar mass eigenstates $\varphi=h,H,A$ can then be written as $g_{\varphi f} = \kappa_{\varphi f} m_f/v$ with the $\kappa_{\varphi f}$ given in Table \ref{tab:kappas}, in which
\begin{table}[b]
\begin{tabular}{|c | c|}
\hline
Type I & Type II\\[2pt]
\hline\hline&\\[-.3cm]
 $\begin{aligned}
\kappa_{hu}&=\kappa_{hd}=\kappa_{h\ell}= -s_{\beta-\alpha}-\frac{c_{\beta-\alpha}}{t_\beta}\\[-3pt]
\kappa_{Hu}&=\kappa_{Hd}=\kappa_{H\ell}= \frac{s_{\beta-\alpha}}{t_\beta}-c_{\beta-\alpha}\\[-3pt]
\kappa_{Au}&=-\kappa_{Ad}=-\kappa_{A\ell}=\frac{1}{t_\beta}
\end{aligned} $
&
$ \begin{aligned}
\kappa_{hu}&=-s_{\beta-\alpha}-\frac{c_{\beta-\alpha}}{t_\beta}\,,\hspace{.95cm}\kappa_{hd}=\kappa_{h\ell}= c_{\beta-\alpha}t_\beta-s_{\beta-\alpha}\\[-3pt]
\kappa_{Hu}&=\frac{s_{\beta-\alpha}}{t_\beta}-c_{\beta-\alpha}\,,\qquad \kappa_{Hd}=\kappa_{H\ell}= -c_{\beta-\alpha}-s_{\beta-\alpha}t_\beta\\[-3pt]
\kappa_{Au}&=\frac{1}{t_\beta}\,,\hspace{2.3cm} \kappa_{Ad}=\kappa_{A\ell}= t_\beta\\[.3cm]
\end{aligned} $\\
\hline
\end{tabular}
\caption{\label{tab:kappas} Couplings between the scalars $h, A, H$ and $H^\pm$ to fermions for 2HDMs of type I and type II.}
\end{table}
the Yukawa couplings correspond to the ones of a two Higgs doublet model of type II (as given in \eqref{eq:2HDMEFT}), or to a two Higgs doublet model of type I, which follow from the Yukawa couplings in \eqref{eq:2HDMEFT} with the replacements $H_2\rightarrow \tilde H_1$. The couplings of the neutral scalars to Dark Matter are given by 
\begin{align}\label{eq:DMcouplings}
g_{h\chi}&=\bigg(\frac{2(c_{\beta-\alpha}+t_\beta s_{\beta-\alpha})}{1+t_\beta^2}-c_{\beta-\alpha}\bigg) \text{Re}[C_\chi]\,,\\
g_{h5}&=\bigg(\frac{2(c_{\beta-\alpha}+t_\beta s_{\beta-\alpha})}{1+t_\beta^2}-c_{\beta-\alpha}\bigg) \text{Im}[ C_5]\,,\\
g_{H\chi}&=\bigg(\frac{2(c_{\beta-\alpha}t_\beta -s_{\beta-\alpha})}{1+t_\beta^2}+s_{\beta-\alpha}\bigg) \text{Re}[C_\chi]\,,\\
g_{H5}&=\bigg(\frac{2(c_{\beta-\alpha}t_\beta -s_{\beta-\alpha})}{1+t_\beta^2}+s_{\beta-\alpha}\bigg) \text{Im}[C_5]\,,\\
g_{A\chi}&= \text{Im}[C_\chi]\,, \qquad g_{A5}=\text{Re}[C_5]\,,
\end{align}
in which we defined the dimensionless couplings $C_\chi=c_\chi v/\Lambda$ and $C_5=c_5 v/\Lambda$. The couplings of the charged Higgs to SM fermions follow from \eqref{eq:2HDMEFT} and read
\begin{align}
\mathcal{L}=-\frac{\sqrt{2}}{v}H^+\sum_{i,j=1}^3\left( \bar u_i \left(\kappa_{H^+d}V_{ij} m_{d_j} P_R- \kappa_{H^+u}m_{u_i}V_{ij}P_L\right) d_j+\kappa_{H^+\ell}\bar \nu m_\ell P_R\ell\right) +h.c.\,,
\end{align}
with $\kappa_{H^+f}=\kappa_{Af}$ for all SM fermions $f$, and $V_{ij}$ are the elements of the CKM matrix.\\

\begin{figure}[t]
\begin{center}
\includegraphics[width=1\textwidth]{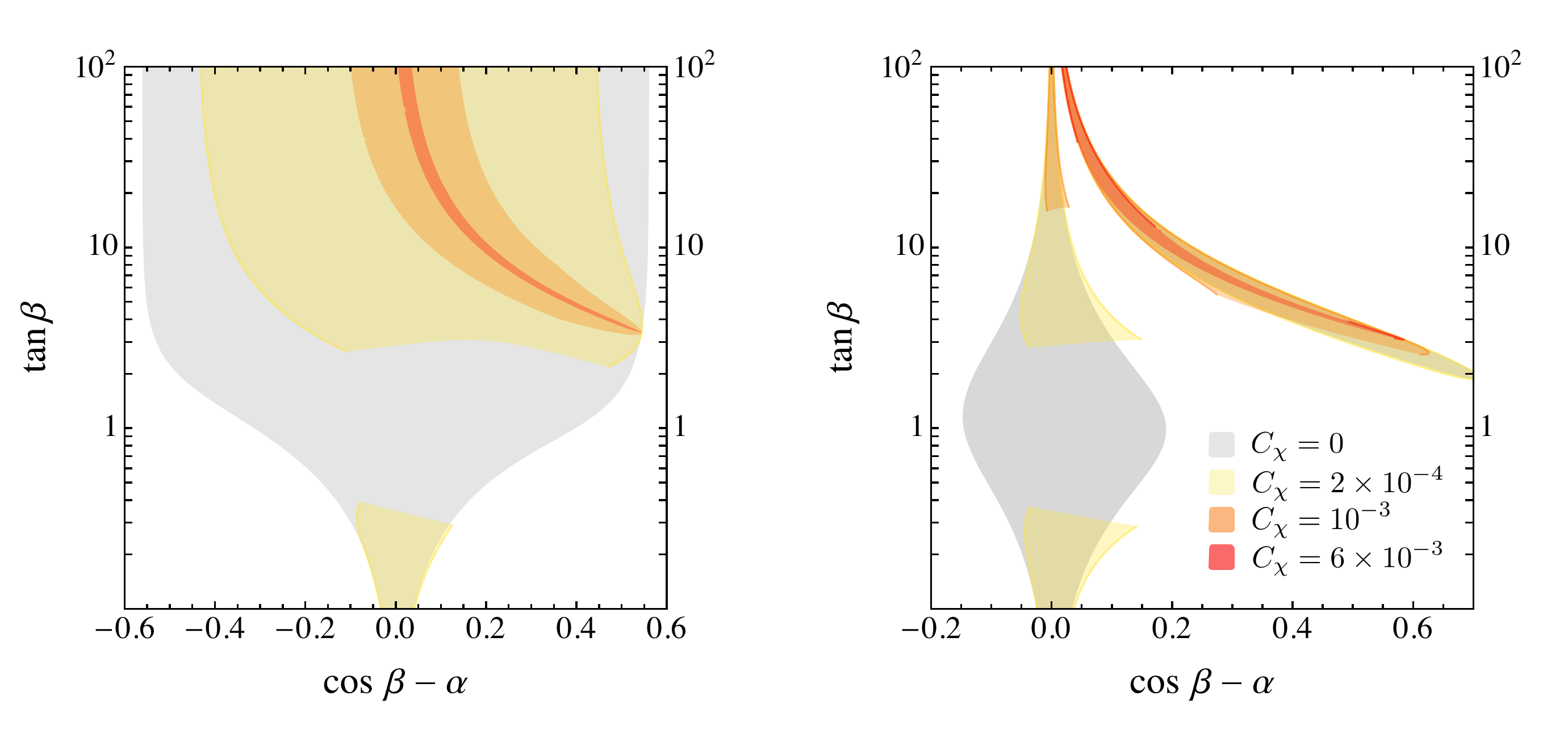} 
\caption{\label{fig:higgsfit}Region in the $\cos (\beta-\alpha)-\tan \beta$ parameter plane of a two Higgs doublet model of type I (left) and type II (right), allowed by a global fit to Higgs signal strength measurements for $C_\chi=0$  (gray), $C_\chi=2\times 10^{-4}$ (yellow), $C_\chi=10^{-3}$ (orange ) and $C_\chi=6\times 10^{-3}$ (red).  }
\end{center}
\end{figure}
Measurements of the Higgs coupling strength in several channels put strong constraints on any possible mixing of the Higgs with new scalar degrees of freedom.\footnote{For simplified models in which the Higgs mixes with scalar mediators that couple to dark matter, measurements of Higgs couplings provide a stronger bound on the mixing angle than any mono-X search \cite{Bell:2016ekl}. } In Figure \ref{fig:higgsfit}, we present a global fit to SM Higgs signal strength measurements
\begin{align}\label{eq:signalstrength}
\mu_X=\frac{\sigma_\mathrm{prod}}{\sigma_\text{prod}^\text{SM}}\frac{\Gamma(h\rightarrow X)}{\Gamma(h\rightarrow X)^\text{SM}}\frac{\Gamma_{h}^\text{SM}}{\Gamma_h}\,,
\end{align}
based on the combination of CMS and ATLAS signal strength measurements presented in \cite{Khachatryan:2016vau}.
Here, $\sigma_\mathrm{prod}$ denotes the production cross section, $\Gamma(h\rightarrow X)$ the partial decay width of the Higgs into the final state $X$ and $\Gamma_h$ the total width of the Higgs. The SM predictions are denoted by the superscript $\text{SM}$. An additional constraint arises from the bound on invisible Higgs decays $\text{Br}(h \rightarrow \text{invisible})<0.23$ \cite{Aad:2015pla,Khachatryan:2016whc}. We consider the generic scenario of a two Higgs doublet model of type I (left panel) and type II (right panel) for which no couplings to dark matter are present for $C_{\chi}=0$. The allowed parameter space for this case is shaded gray. We further show the global fit for three additional values of $C_{\chi}=2\times 10^{-4}, 10^{-3}, 6\times 10^{-3}$ with the respective parameter space allowed by all constraints shaded yellow, orange and red. The Dark Matter mass has been fixed to $m_\chi=0$. The parameter $\text{Im} [C_5]$ also allows for Higgs couplings to Dark Matter, but leads to the same results, up to a weaker sensitivity on the Dark Matter mass in the case of the pseudoscalar coupling. The parameter space that survives for large values of $\text{Re} [C_\chi]$ or $\text{Im} [C_5]$ corresponds to the region in which $g_{h\chi}=g_{h5}=0$. This parameter space is not stable under additional contributions from loop-induced Higgs couplings or additional operators, such as $H_iH_i^\dagger \bar \chi \chi$, $i=1,2$.
It follows therefore that either the Wilson coefficients $\text{Re} [C_\chi]$ and $\text{Im} [C_5]$ need to be severely suppressed, or the Higgs decays need to be kinematically disallowed. Other scenarios are excluded by Higgs coupling strength measurements even in the decoupling limit. 
\subsection{Flavour and  Electroweak Precision Observables}
\begin{figure}[t]
\begin{center}
\includegraphics[width=1\textwidth]{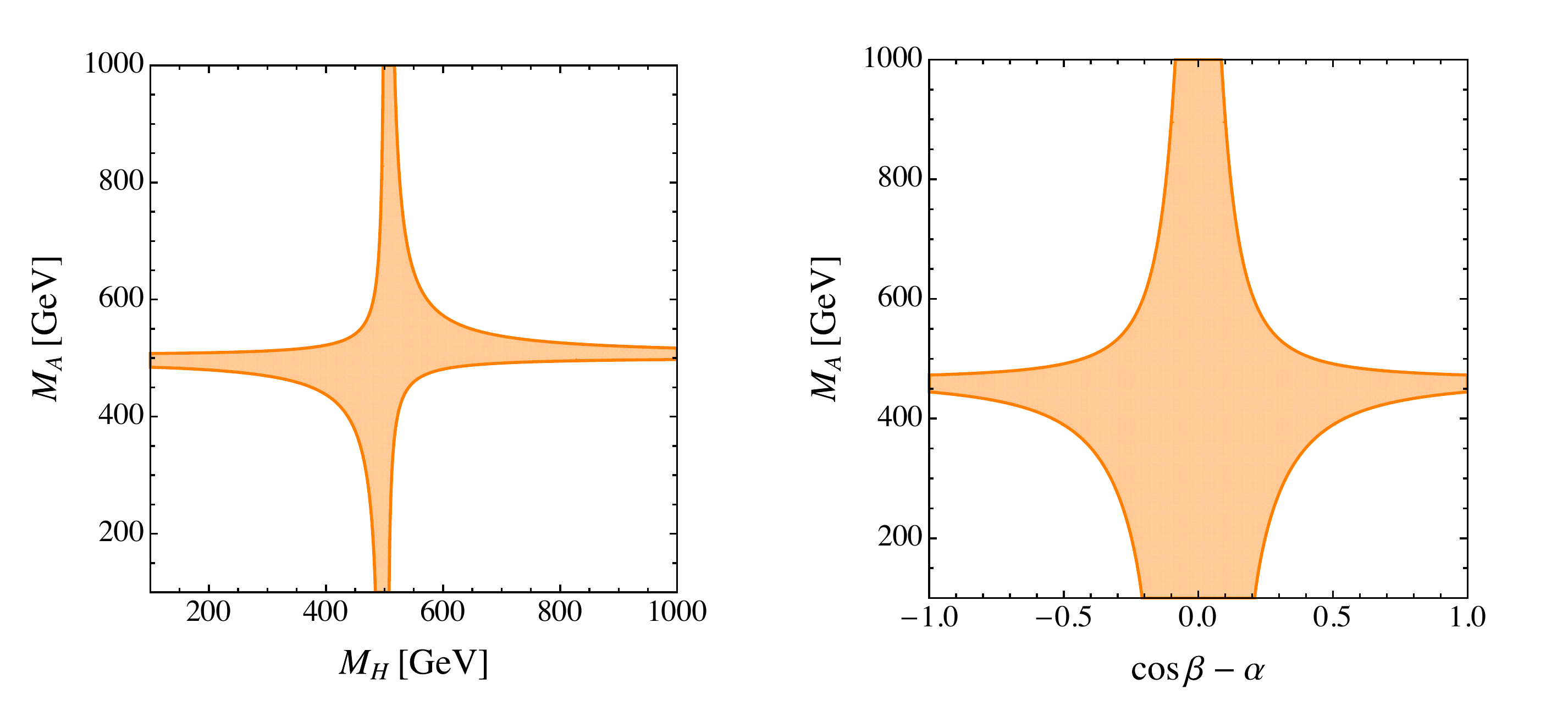} \vspace{-.6cm}
\caption{\label{fig:ewpt} Left: Parameter space allowed by a combined fit to the oblique parameters at the $95\%$ C.L. in the $M_H-M_A$ plane for $M_{H^+}=500$ GeV and $\cos(\beta-\alpha)=0$. Right: Parameter space allowed by a combined fit to the oblique parameters at the $95\%$ C.L. in the $\cos(\beta-\alpha) - M_A$ plane for fixed masses of $M_H=M_{H^+}=500$ GeV. }
\end{center}
\end{figure}
Natural flavour conservation ensures the absence of tree-level flavour changing vertices of the neutral spin-0 particles $h, H$ and $A$. Contributions to flavour changing neutral currents (FCNCs) at one loop from charged Higgs exchange therefore lead to the strongest bounds on the model parameters. In particular, measurements of $b \rightarrow s \gamma$ decays based on the Belle dataset \cite{Belle:2016ufb} require at 95$\%$ C.L.,  $M_{H^\pm}> 569 - 795$ GeV for two Higgs doublet modes (2HDMs) of type II and $M_{H^\pm}> 268 - 504$ GeV for type I Yukawa couplings and $\tan \beta =1$, where the range depends on the method applied to derive that bound  \cite{Misiak:2017bgg}. While this constraint is rather independent from $\tan \beta$ for type II 2HDMs, it scales like $\propto 1/\tan^2 \beta$ in the case of type I 2HDMs. As a consequence, for $\tan \beta>2$, flavour constraints become less important than collider searches for the latter case. Anticipating the unitarity and perturbativity bounds derived below, large values of $\tan \beta$ are strongly disfavoured even for Yukawa sectors of type I 2HDMs and we adopt the constraint $M_{H^\pm}> 500$ GeV in the following. Additional model-independent constraints arise from corrections to $B_s-\bar B_s$ meson mixing and to $Z\to b\bar b$ decays from charged Higgs loops. For $M_{H^\pm}=500$ GeV, these contributions lead to the constraint $\tan \beta > 0.9$ \cite{Geng:1988bq,Deschamps:2009rh}. It should be stressed that indirect bounds are subject to change if more complete models are considered and contributions from additional particles to the relevant observables are taken into account.

\begin{figure}[t]
\begin{center}
\includegraphics[width=1.\textwidth]{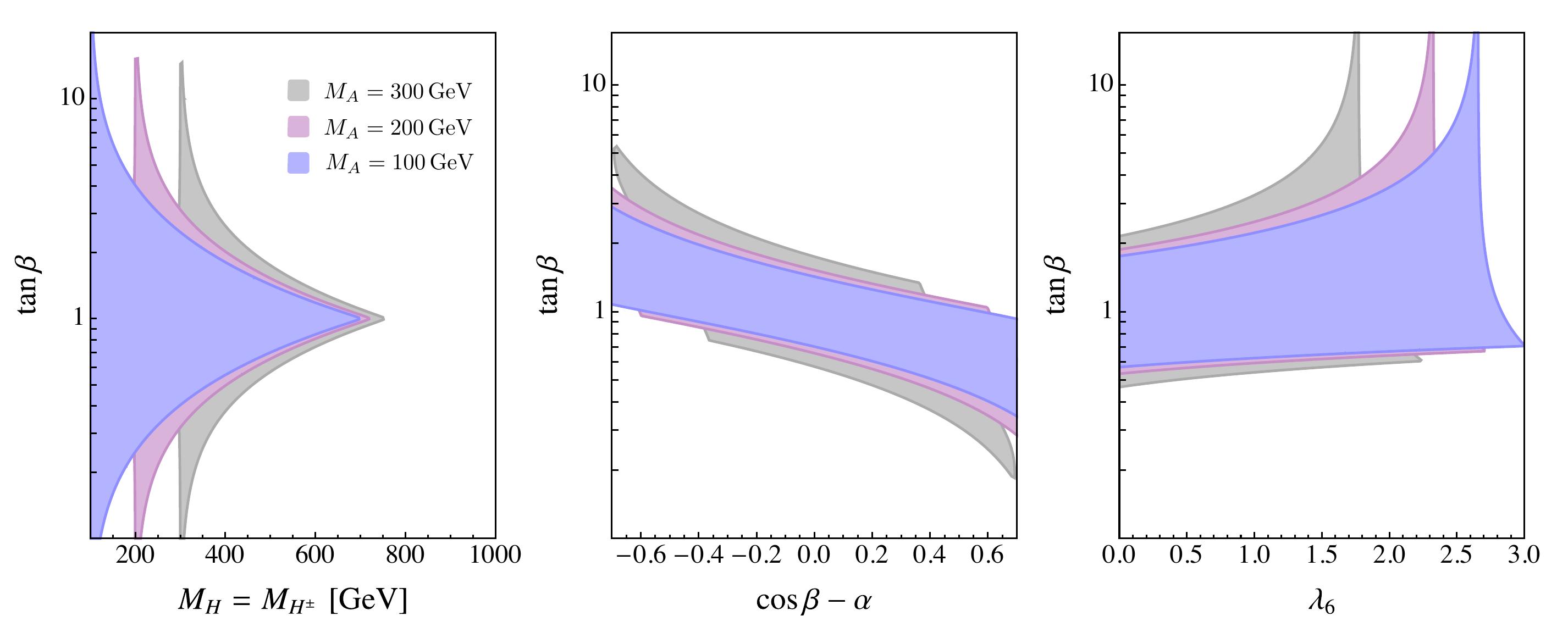} 
\caption{\label{fig:UPS} Left: Parameter space allowed by stability, unitarity and perturbativity constraints for three different values of pseudoscalar masses $M_A=100$ GeV (blue), $200$ GeV (purple) and $300$ GeV (gray), $M_H=M_H{\pm}$, and $\cos(\beta-\alpha)$=0. Center: Parameter space allowed by stability, unitarity and perturbativity constraints in the $\cos(\beta-\alpha)-\tan\beta$ plane for $M_{H}=M_{H^\pm}=500$ GeV. Right: The effect of a non-vanishing quartic coupling $\lambda_6$ on the parameter space in the alignment limit $\cos(\beta-\alpha)=0$.
 }
\end{center}
\end{figure}

At the one-loop level, the neutral and charged scalars and the pseudoscalar modify electroweak precision parameters, such as the tree-level relation between $Z$ and $W^\pm$ boson masses set by electroweak symmetry breaking. These effects are independent from $\tan \beta$, because the couplings of the scalars and the pseudoscalar to gauge bosons only depend on $\cos(\beta-\alpha)$. The corresponding constraints are therefore valid for both type I and type II 2HDMs and constrain the mass splittings between the heavy spin-0 mass eigenstates $M_H, M_A$ and $M_{H^+}$ and the mixing angle $\cos(\beta-\alpha)$. Taken into account the preference for the alignment limit $\cos(\beta-\alpha)=0$ of the global fit to Higgs signal strength measurements, and flavour constraints, we show the allowed parameter space by a 95$\%$ C.L. fit to the oblique parameters $S, T$ and $U$ in the $M_A-M_H$ plane for fixed $M_{H^\pm}=500$ GeV and $\cos(\beta-\alpha)=0$ on the left panel of Figure~\ref{fig:ewpt}. A clear preference for almost degenerated masses $M_H\approx M_{H^\pm}$ or $M_H\approx M_{A}$ is evident. This can be understood by the restoration of the global custodial symmetry present in the SM Higgs potential in the full 2HDM Higgs potential in these limits \cite{Gerard:2007kn, Grzadkowski:2010dj}. Since we are interested in scanning the range of pseudoscalar mediators, we choose $M_H=M_{H^\pm} = 500$ GeV and present the allowed parameter space in the $\cos(\beta-\alpha)-M_{A}$ plane. Apart from a fully degenerate spectrum $M_A\approx M_H\approx M_{H^\pm}$, electroweak precision constraints prefer the alignment limit and in the case of 2HDMs of type I result in a stronger constraint on $\cos(\beta-\alpha)$ than the global fit to Higgs coupling strength measurements for $\tan\beta \gtrsim 1$.
As in the case of flavour observables, it should be stressed that the constraints from electroweak precision observables are indirect and sensitive to the presence of additional particles charged under $SU(2)_L\times U(1)_Y$, which can lead to cancellations in complete models. The bounds presented here should therefore only serve as a guide.

\begin{figure}[t]
\begin{center}
\includegraphics[width=1\textwidth]{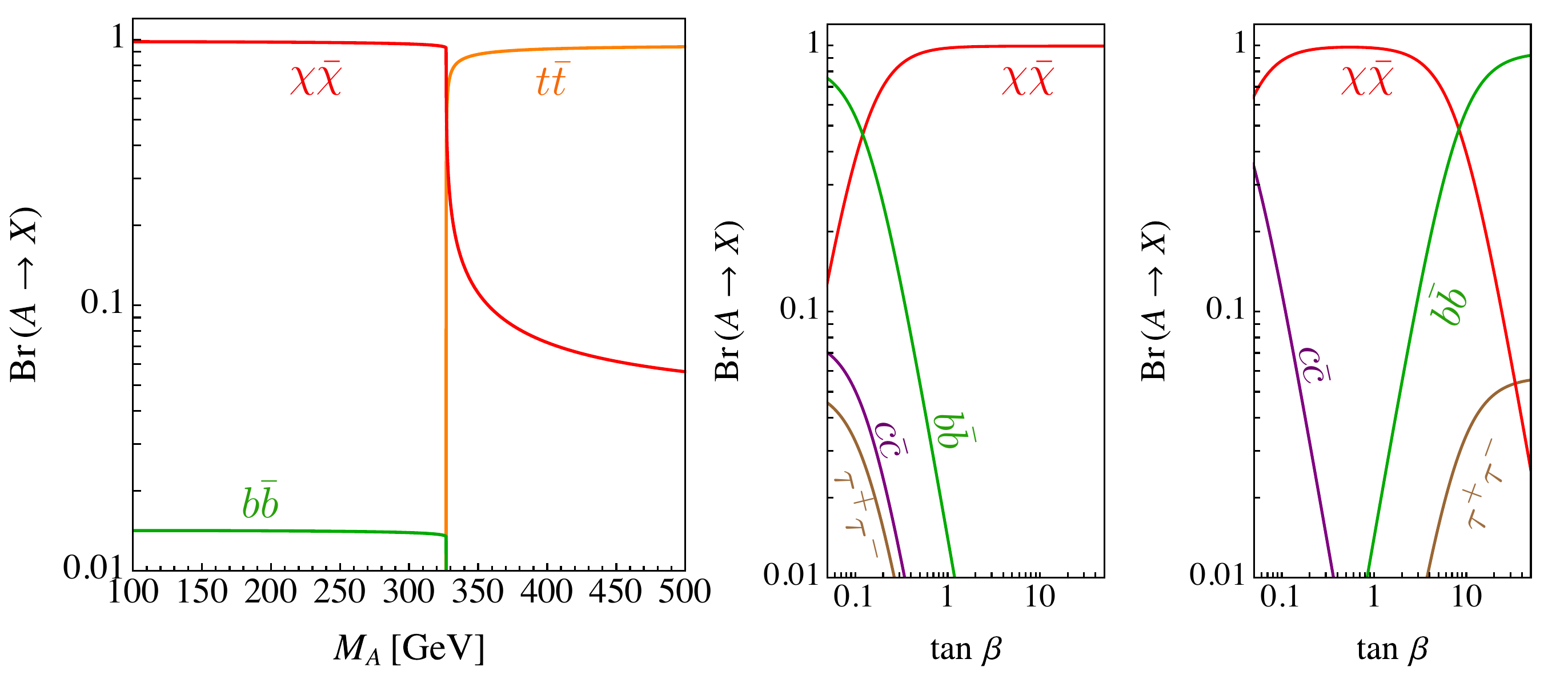}
\caption{\label{fig:ABRs} Left: The dominant branching ratios of the pseudoscalar $A$ in dependence of the pseudoscalar mass $M_A$ for $M_H=M_H^\pm=500$ GeV, $\tan\beta=1$ and $\cos(\beta-\alpha)=0$. Right: The dominant branching ratios of the pseudoscalar $A$ in dependence of $\tan\beta$ for the pseudoscalar mass $M_A=200$ GeV, $M_H=M_H^\pm=500$ GeV and $\cos(\beta-\alpha)=0$.}
\end{center}
\end{figure}

\subsection{Unitarity, Perturbativity and Stability Requirements}
\label{sec:stab}

Stability of the scalar potential \eqref{eq:VH} requires that the quartic couplings fulfill the following conditions \cite{Gunion:2002zf}
\begin{align}
\lambda_1>0\,,\qquad \lambda_2>0\,, \qquad \lambda_3>-\sqrt{\lambda_1\lambda_2}\qquad \lambda_4+\lambda_3 +\sqrt{\lambda_1\lambda_2}>0\,.
\end{align}
Further, perturbativity constraints on the separate quartic couplings require $|\lambda_i|< 4\pi$ $i=1,\ldots, 4 $. 
Partial wave unitarity translates in the condition that the eigenvalues of the relevant submatrices of the scattering matrix have eigenvalues $s_{i}$ with $|s_i|< 8 \pi$ for all $i$ \cite{Ginzburg:2005dt, Eriksson:2009ws}. Given that the potential \eqref{eq:VH} is completely fixed by the masses of the spin-0 particles $M_h, M_H, M_A, M_{H^\pm}$ and the mixing angles $\tan\beta$ and $\cos (\beta-\alpha)$, stability, perturbativity and unitarity requirements lead to strong constraints on the quartic couplings. In particular, a large mass splitting $M_A< M_H, M_{H^\pm}$ requires sizable quartics and is therefore constrained by perturbativity and unitarity. This is illustrated in the left panel of Figure~\ref{fig:UPS} for three different values of pseudoscalar masses $M_A=100$ GeV (blue), $200$ GeV (purple) and $300$ GeV (gray). For a fixed, sizable mass splitting only a small range of values for $\tan\beta$ are allowed. In the center panel of Figure~\ref{fig:UPS}, we show the allowed parameter space in the $\cos(\beta-\alpha)-\tan\beta$ plane for $M_{H}=M_{H^\pm}=500$ GeV. Taking into account the constraint from electroweak precision observables for these masses $|\cos(\beta-\alpha)|\lesssim 0.2$, results in a constraint $0.5\lesssim\tan\beta\lesssim 2.5$. We note that this constraint can be considerably relaxed in more general models which allow for additional quartic couplings. As an example, we show the effect of adding the quartic coupling $\Delta V_H = \lambda_6 H_1^\dagger H_1 \,H_1H_2^\dagger +h.c.$ to the potential \eqref{eq:VH} with real values $\lambda_6=0-3$ in the right panel of Figure~\ref{fig:UPS}. In this case, larger values are possible, but still disfavoured with respect to smaller values of $\tan\beta=\mathcal{O}(1)$. \\
Additional perturbativity constraints can be derived for the Yukawa couplings in \eqref{eq:2HDMEFT}. In particular the top Yukawa coupling becomes non-perturbative for $\tan\beta\lesssim 0.3$ for both type I and type II 2HDMs \cite{Branco:2011iw}. This constraint is automatically fulfilled once the stability, perturbativity and unitarity constraints on the scalar potential are taken into account.

\begin{figure}[t]
\begin{center}
\includegraphics[width=1\textwidth]{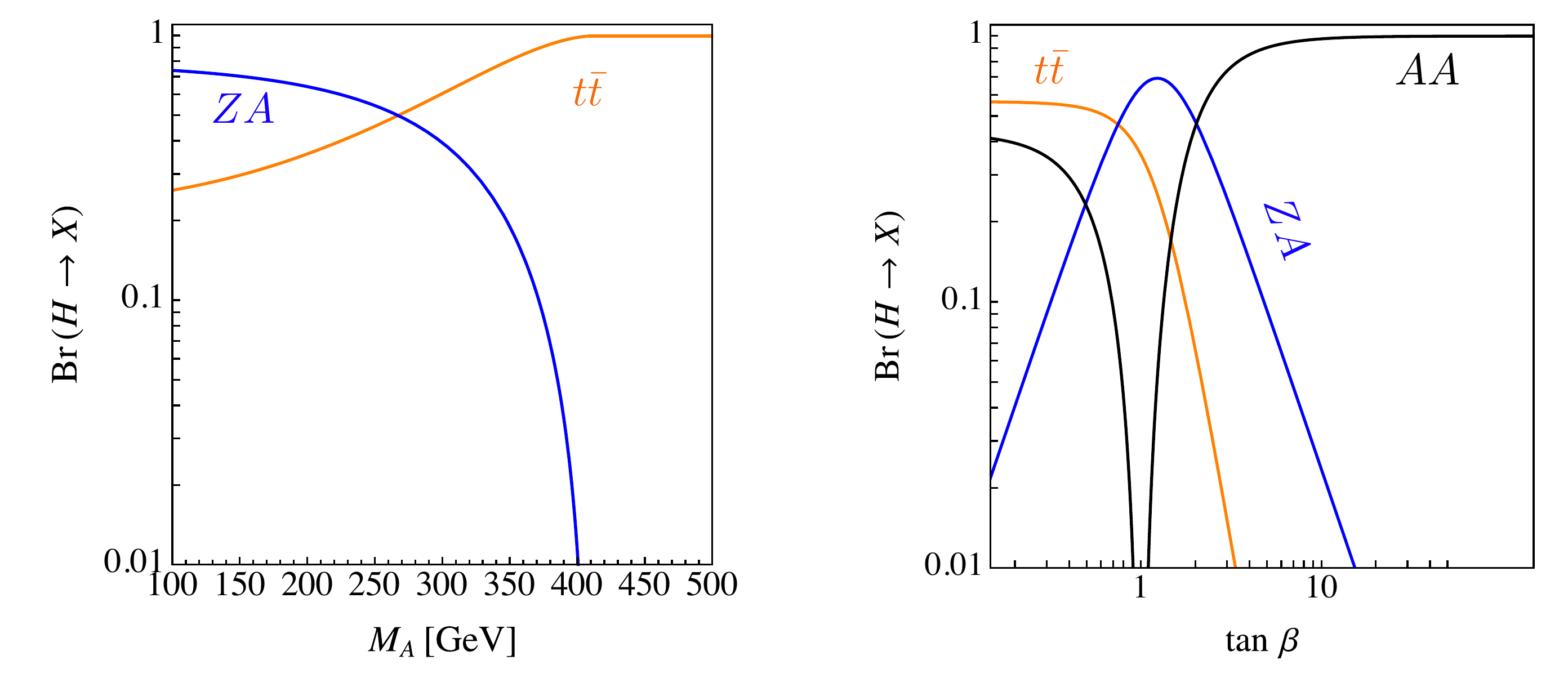}
\caption{\label{fig:HBRs} Left: The dominant branching ratios of the heavy scalar $H$ in dependence of the pseudoscalar mass $M_A$ for $M_H=M_H^\pm=500$ GeV, $\tan\beta=1$ and $\cos(\beta-\alpha)=0$. Center (and right): The dominant branching ratios of the heavy scalar $H$ in dependence of $\tan\beta$ for the pseudoscalar mass $M_A=200$ GeV, $M_H=M_H^\pm=500$ GeV and $\cos(\beta-\alpha)=0$ and the Yukawa sector of a 2HMD of type I (II).}
\end{center}
\end{figure}

\subsection{Collider Searches}

\begin{figure}[t]
\begin{center}
\includegraphics[width=1\textwidth]{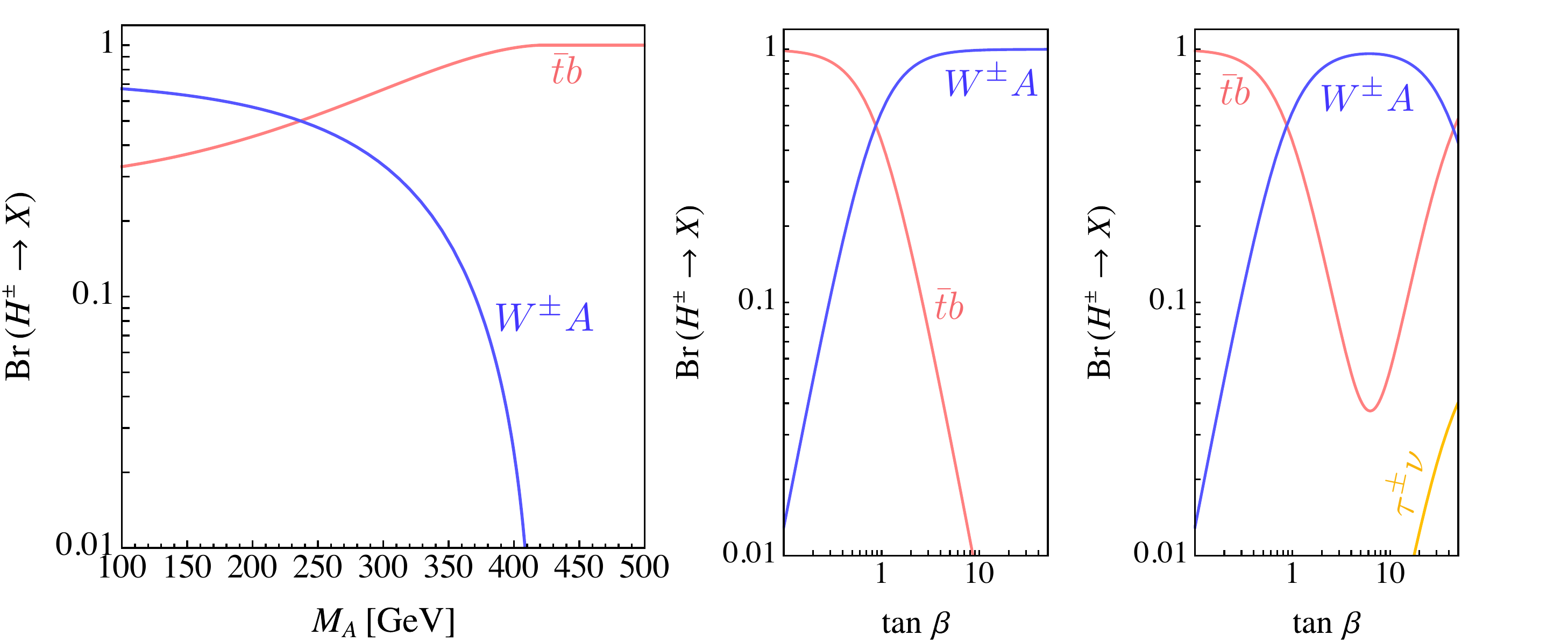}
\caption{\label{fig:HpmBRs} Left: The dominant branching ratios of the charged scalar $H^\pm$ in dependence of the pseudoscalar mass $M_A$ for $M_H=M_{H^\pm}=500$ GeV, $\tan\beta=1$ and $\cos(\beta-\alpha)=0$. Center (and right): The dominant branching ratios of the charged scalar $H^\pm$ in dependence of $\tan\beta$ for the pseudoscalar mass $M_A=200$ GeV, $M_H=M_H^\pm=500$ GeV and $\cos(\beta-\alpha)=0$ with Yukawa couplings as in a 2HDM of type I (II).}
\end{center}
\end{figure}
Collider searches for the heavy resonances $A, H$ and $H^\pm $ directly constrain their masses and couplings to SM particles. We consider only the alignment limit $\cos(\beta-\alpha)=0$, preferred by Higgs and electroweak precision bounds. In this case, for the pseudoscalar $A$, only couplings to fermions are relevant. The left panel of Figure \ref{fig:ABRs} shows the scaling of the branching ratios of $A$ for different values of $M_A$ and $M_H=M_{H^\pm}=500$ GeV, $c_5=1, c_\chi=0$, $m_\chi=1$ GeV and $\tan \beta=1$. Branching ratios not shown in this plot are smaller than $1\%$. For masses below the top threshold, $M_A<2\,m_t$, the branching ratio into Dark Matter dominates, as expected from the relative coupling strength of the coupling to Dark Matter and to b quarks $g_{Ab}/g_{A\chi}\sim m_b\Lambda/v^2$. The center and right panel show the variation of the branching fractions with $\tan \beta$ for fixed $M_A=200$ GeV and a Yukawa sector as in a 2HDM of type I (center) and type II (right). In both types of models, the branching ratio into Dark Matter dominates for $0.2 \lesssim \tan \beta \lesssim 8$. For larger values of $\tan\beta$, decays into $b\bar b$ pairs become the dominant decay channel in a type II 2HDM. The three-body decays are not shown in Figure~\ref{fig:ABRs}, although the branching ratio $A \to t \bar t^*\to W^- t \bar b$ can become non-negligible in parts of the parameter space close to the top-threshold. \\ 
In the case of the heavy neutral scalar $H$, the branching ratios are shown in Figure \ref{fig:HBRs}, for the parameters $M_H=M_{H^\pm}=500$ GeV, $c_5=1, c_\chi=0$, $m_\chi=1$ GeV. In the left panel, we further fix $\tan \beta =1$ and show the dependence of the branching ratios on $M_A$. 
In the right panel, we instead fix $M_A=200$ GeV and vary $\tan\beta$. The result holds for both 2HDMs of type I and II, because the dominant branching ratios are $\text{Br}(H\to \bar t t)$, $\text{Br}(H \to AA)$ and $\text{Br}(H\to ZA)$. The $\text{Br}(H\to ZA)$ is the most important decay channel of the heavy scalar for $3\geq\tan\beta\geq 1$ and gives rise to a mono-$Z$ final state for the dominant pseudoscalar decay channel $A\to \bar \chi \chi$. If $M_H \leq 2 M_A$ and $M_H> M_A + M_Z$, the decay $H \to AA$ is not kinematically allowed. In this case, $\text{Br}(H \to ZA)$ is the dominant branching ratio also for larger values on $\tan\beta$. It is intriguing that the parameter space giving rise to a mono-$Z$ signal is in agreements with the bounds discussed in the previous sections. \\
For the heavy charged scalar $H^\pm$ the dependence of the branching ratios on $M_A$ is shown on the left panel of Figure \ref{fig:HpmBRs} for  $M_H=M_{H^\pm}=500$ GeV, $c_5=1, c_\chi=0$, $m_\chi=1$ GeV and $\tan \beta=1$. The center and right panel of Figure \ref{fig:HpmBRs} show the dependence of the branching ratios of the charged scalar on $\tan \beta$ for fixed $M_A=200$ GeV and the Yukawa couplings as in a 2HDM of type I (center) and type II (right), respectively. For $1 \lesssim \tan \beta \lesssim 13$ and $M_{H^\pm}> M_A+M_{W^{\pm}}$, the charged scalar dominantly decays into the pseudoscalar and a $W^\pm$ boson, resulting in a mono-$W^\pm$ signature for the dominant decay mode of the pseudoscalar $A\to \bar \chi\chi$. In contrast to the $H\to A Z$ channel, the branching ratio $\text{Br}(H^\pm\to W^\pm A)$ remains large for values of $\tan \beta =10$ for both Yukawa sectors of type I and type II and also for larger values in the case of type I. Analytic expressions for the corresponding partial decay widths of the pseudoscalar, heavy scalar and charged scalar are collected in Appendix \ref{sec:pw}. \\
Pseudoscalars with masses below the top mass threshold can be constrained by searches for $pp\to A\, b\bar b\to \tau^+\tau^-\bar b b$ \cite{Sirunyan:2017uvf} and $A\to Zh$ \cite{ATLAS:2017nxi, Sirunyan:2017wto}, but the former decay is strongly suppressed since $\text{Br}(A\to \tau^+\tau^-)< 1\%$, while the latter is not allowed in the decoupling limit. Collider searches for heavy scalar resonances in $H \to t\bar t$ lead to the constraints $\tan\beta\gtrsim 1$ for $M_H =500$ GeV \cite{ATLAS:2016pyq}. Searches for charged scalars are most sensitive in the $H^+ \to \tau^+ \nu$ final state \cite{Aaboud:2016dig, CMS:2016szv}, where the corresponding branching ratio is very small if the pseudoscalar is light enough to enable the $H^\pm \to W^\pm A$ decay. In a recent analysis, ATLAS has obtained limits for the $H^+ \to t\bar b$ decay, for which the branching ratio can be sizable in our model \cite{ATLAS:2016qiq}. Both searches put no relevant constraints on $\tan \beta$ for $M_{H^\pm}=500$ GeV \cite{Arbey:2017gmh}.

\newpage

\section{Constraints from Direct Detection, Indirect Detection and the Relic Abundance of Dark Matter}\label{sec:constraints2}

The constraints discussed in the previous section are relevant to all two Higgs doublet models with an additional coupling to fermions, and further constraints arise if the fermions $\chi$ constitute some or all of the Dark Matter. In the following, we discuss bounds on the parameter space of the model presented in \eqref{eq:2HDMEFT} from searches for Dark Matter with direct and indirect detection experiments and from the measurement of the relic abundance of Dark Matter in the case that $\chi$ is the only Dark Matter candidate. We note that these constraints are model dependent and subject to change in more complicated models that can be described in the appropriate limit by the EFT \eqref{eq:2HDMEFT}. \\

The relic abundance of dark matter has been precisely measured by the Planck collaboration $(\Omega_{\chi}h^{2})_\text{Planck} = 0.1198 \pm 0.0015$ \cite{Ade:2015xua}. Taking this measurement at face value would fix the relation between the mass of the mediator, the dark matter mass and the coupling strength. This relation can be misleading regarding the allowed parameter space if the model from which it is derived is incomplete as is explicitly the case for simplified models such as the one discussed here. For example, if the predicted annihilation is too effective, the resulting under-abundance can be explained by the presence of a second stable particle. On the other hand, if the particle the mediator mainly couples to is stable on collider scales, but eventually decays into a lighter, stable species of dark matter, collider searches for mono-$X$ signatures could discover such a dark sector, even if the 
prediction for the annihilation cross section would suggest an over-closure of the universe.  In anticipation of the results of the discussion in this section, we note that it is however remarkable that the parameter space for which we can recover the observed relic density of Dark Matter of the model we discuss, is in agreement with the bounds derived in the previous section that are independent of the dark sector.    \\

\begin{figure}	\centering
	\includegraphics[width= 1.05\textwidth]{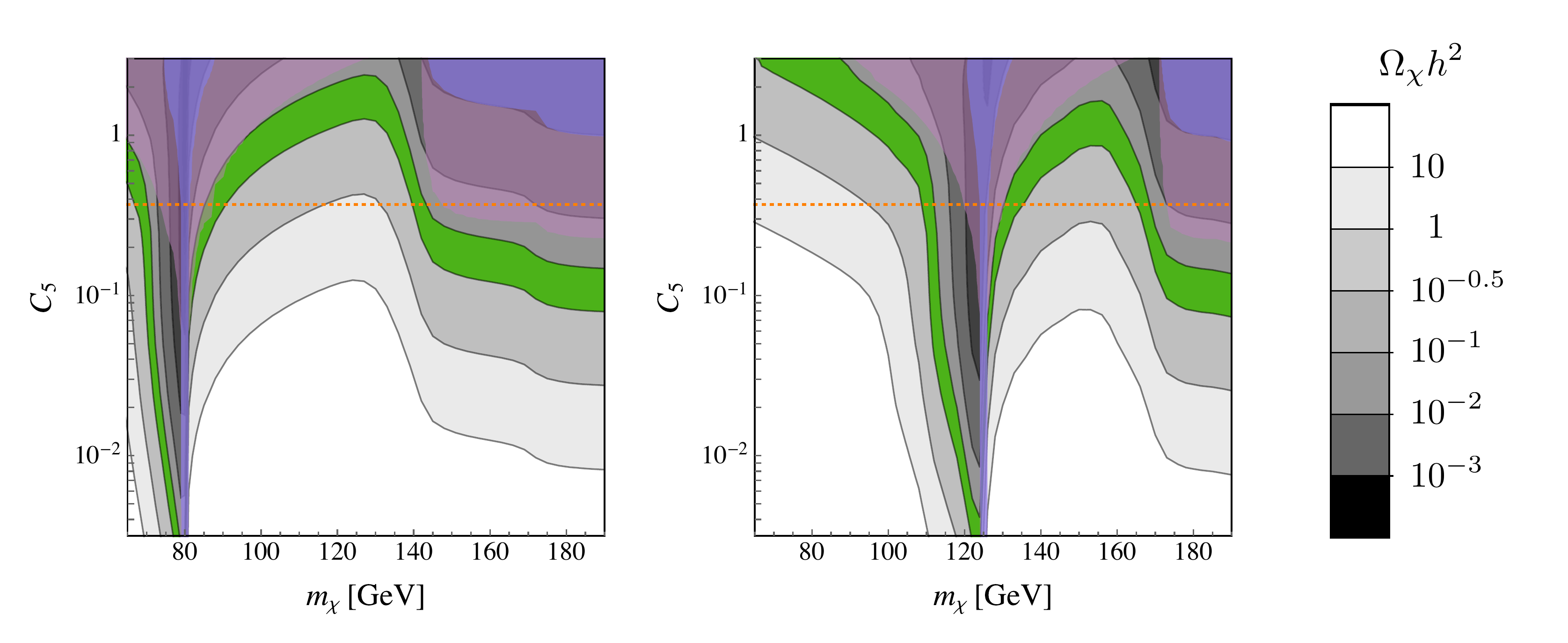}	\caption{The left (right) panel shows contours of the predicted Dark Matter annihilation cross section in units of $\Omega_\chi\,h^2$ for $\tan\beta=1$ and a mediator mass of $M_{A}$=160 (250) GeV in dependence of $m_\chi$ and $C_5$. The green region corresponds to $0.04> \Omega_{\chi} h^{2} > 0.13$, and the purple region is excluded by the CMB-measurement from Planck. The blue shaded region shows the projected sensitivity of a future CTA measurement, and the dashed orange line corresponds to the value chosen for the analyses in the remainder of the paper.
}
	\label{fig:relicchi}
\end{figure}

In order to compute the annihilation cross section for the dark matter candidate $\chi$ into SM particles, we use MicrOmegas version 4.3.1 \cite{Belanger:2013oya}. We show the prediction for the relic abundance $\Omega_\chi h^2$ for $M_H=M_{H^{\pm}}=500$ GeV, $\cos(\beta-\alpha)=0$, $C_\chi=0$ and $\tan\beta=1$ in the $m_\chi - C_5$ plane in Figure~\ref{fig:relicchi}. In the left and right panel we set $M_A=160$ GeV and $M_A=250$ GeV, respectively. The parameter space for which the relic density is within $0.13> \Omega_{\chi} h^{2} > 0.04$ is shaded green. The shape of these contours can be understood by the resonant enhancement of the annihilation cross section on the $A$-pole at $m_\chi=M_A/2$ and by the annihilation channels  $\chi\bar \chi \to A h$  and $\chi\bar \chi \to A \to t\bar t$  opening up for $m_\chi=(M_A+M_h)/2$ and $m_\chi=m_t$, respectively. We further consider constraints on the annihilation cross section from distortions of the cosmic microwave background (CMB). The latest measurement from Planck translates into the bound \cite{Altmannshofer:2016jzy}
\begin{equation}
	f_\text{eff}\ \frac{(\sigma v)_\text{ann}}{m_{\chi}}\ \lesssim 3\ \times 10^{-28} \frac{\text{cm}^{3}}{\text{s GeV}}\ ,
	\label{eq:CMBbound}
\end{equation}
\begin{figure}	\centering
	\includegraphics[width= 1.05\textwidth]{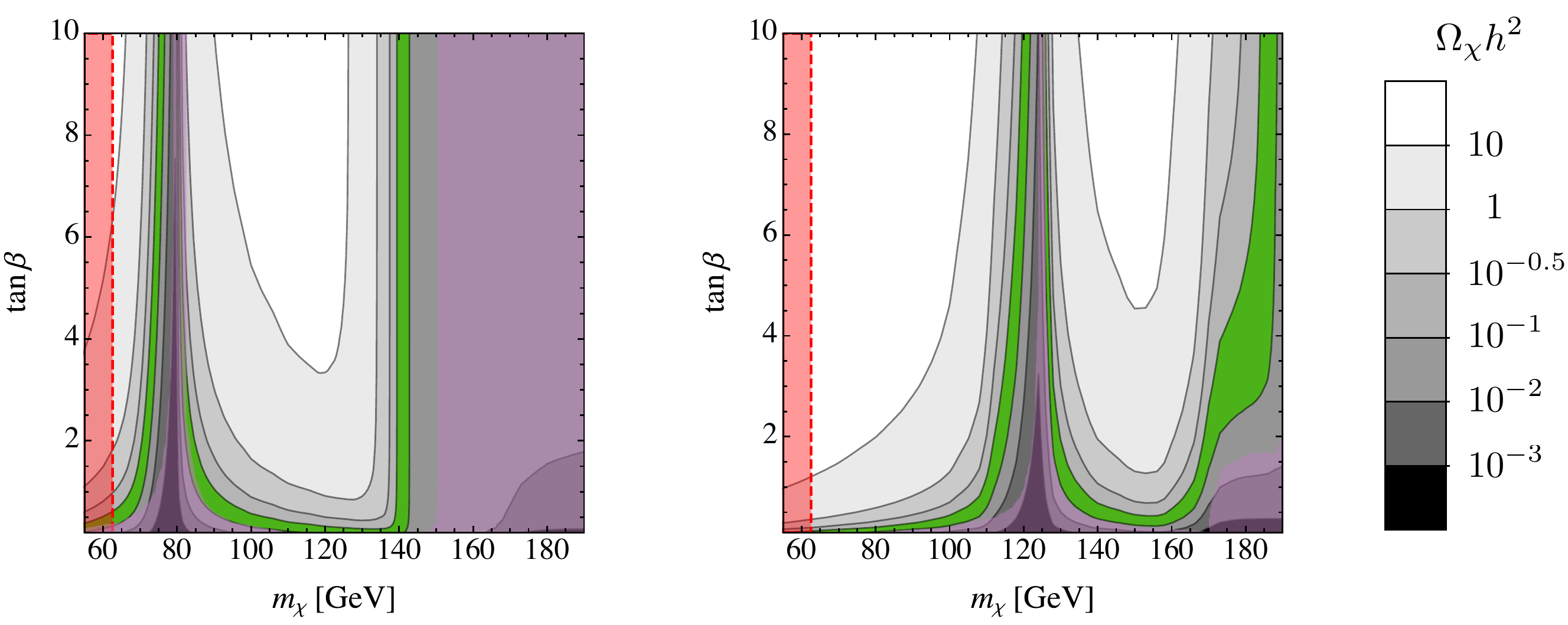}
		\includegraphics[width= 1.05\textwidth]{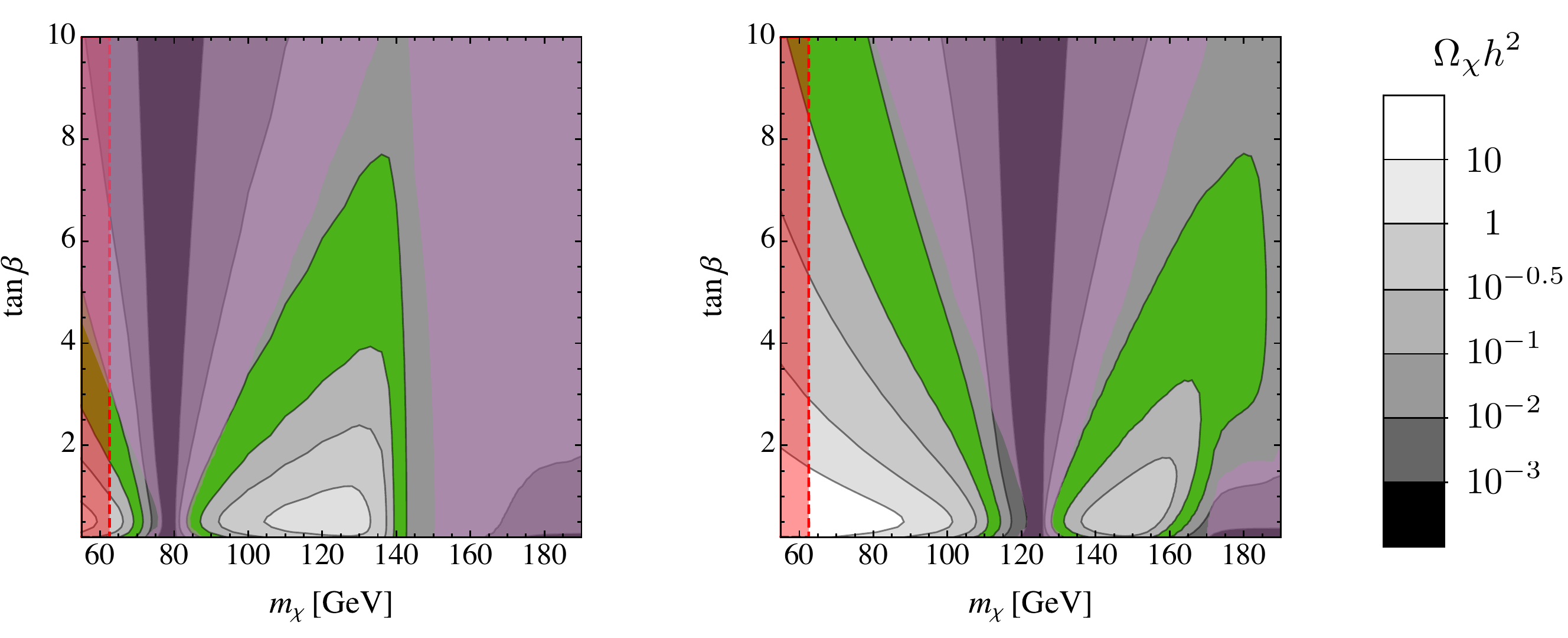}	\caption{The left (right) panel shows contours of the predicted Dark Matter annihilation cross section in units of $\Omega_\chi\,h^2$
		for $C_{5} = 0.37$ and a mediator mass of $M_{A}$=160 (250) GeV in dependence of $m_\chi$ and $\tan \beta$. The contours in the upper (lower) panels correspond to a Yukawa sector of a type I (II) 2HDM. The green region corresponds to $0.13> \Omega_{\chi} h^{2} > 0.04$, the purple region is excluded by the CMB-measurement from Planck, and the red region is disfavoured by Higgs measurements. }
	\label{fig:relictype1}
\end{figure}
in which $f_\text{eff}=0.35$ is the redshift-dependent efficiency
factor evaluated at the time of the last scattering for the dominant final state $\chi\bar\chi \to b\bar b $ throughout most of the parameter space \cite{Slatyer:2009yq}. The parameter space excluded by this constraint is shaded purple in Figure~\ref{fig:relicchi}. 
A weaker constraint is projected for the limits expected by the measurement of the cosmic $\gamma$-ray spectrum by the Cherenkov Telescope Array (CTA) in case of the non-observation of a signal \cite{Balazs:2017hxh}. The corresponding parameter space is shaded blue in Figure~\ref{fig:relicchi}. Both the CMB and CTA constraints do not cut into the parameter space preferred by the relic density measurement. Collider searches for mono-X final states are most sensitive to the parameter space for which the mediator can resonantly decay into Dark Matter $m_\chi < M_A/2$.  In Figure~\ref{fig:relicchi}, we also indicate the benchmark value for $C_5= 0.37$, which corresponds to $c_5=1.5\times 10^{-3}\, (\Lambda/\text{GeV})$, by the dashed orange line.  \\

Constraints from Direct Detection experiments are considerably weaker, because the pseudoscalar mediated Dark Matter-nucleon cross section is suppressed by the non-relativistic Dark Matter velocity. Only purely scalar currents lead to unsuppressed interactions. The currently strongest bound at $m_\chi =  30$ GeV from XENON1T is $\sigma^{\text{XENON1T} }_{\chi-nucleon} \approx 10^{-47}$ cm$^{2}$ and leads to the constraint $C_\chi \lesssim 0.011$ from the exchange of the SM Higgs for the maximal value of $g_{h\chi}$ at $\cos(\beta-\alpha)=0$ for $\tan\beta=1$ \cite{Aprile:2017iyp}.\footnote{The contribution from the exchange of the heavy scalar $H$ to the Dark Matter-nucleus cross section is weaker by a factor $M_h^4/M_H^4$ and further suppressed by $g_{H_\chi}$, which vanishes at $\tan\beta=1$ and $\cos(\beta-\alpha)=0$.} 
\\

The dependence of the annihilation cross section on $\tan\beta$ is illustrated in the plots shown in the upper (lower) panels of Figure~\ref{fig:relictype1} for Yukawa couplings as in a two Higgs doublet model of type I (II). The contours in the left and right panels are again derived with a mediator mass of $M_A=160$ GeV and $M_A=250$ GeV, respectively. We further set $\cos(\beta-\alpha)=0$, $C_5= 0.37$, and $C_\chi= 0$. The parameter space preferred by the measurement of the relic density, $0.13> \Omega_{\chi} h^{2} > 0.04$, is shaded green and the  purple shaded region is excluded by the CMB measurement. We further indicate the region for which the SM Higgs boson can in principle decay into pairs of Dark Matter, which is shaded red in Figure~\ref{fig:relictype1}. If $\text{Re} \, C_\chi = \text{Im} \,C_5= 0$ exactly, this constraint is irrelevant, but even for small values $C_5=\mathcal{O}(10^{-4})$, the global fit to Higgs coupling strength measurements excludes most of the parameter space if $m_\chi < M_h/2$. From Figure~\ref{fig:relictype1}, it follows that values of $10> \tan\beta> 1$ are favoured by the measurement of the relic density for $M_h/2 < m_\chi < M_A/2$ unless the Dark Matter mass is very close to $M_A/2$ in a 2HDM of type I. This parameter space is independently favoured by the constraints derived from perturbativity, unitarity and stability of the scalar potential in Section~\ref{sec:stab}.

\section{Mono-X Searches}\label{sec:collider}

\begin{table}
\begin{tabular}{|c|c|c|}
\hline 
 $\text{jets}+E^\text{miss}_T$ & $Z+E^\text{miss}_T$ & $t\bar{t}+E_T^\text{miss}$\\[2pt] 
\hline \hline
$E_T^{\text{miss}} > 250 $ GeV &$p_T^\ell > 25/20 $ GeV  &  $E_T^{\text{miss}} > 200 $ GeV  \\[2pt] 
$p_T^{j} > 250 $ GeV & $M_Z-15 < m_{\ell\ell} < M_Z + 10 $ GeV   & lepton veto $p_T^{l} > 10 $ GeV \\[2pt] 
$\left| \eta_j \right| < 2.4$ &$\left| \eta_\ell \right| < 2.4$ & jets $\geq 4$ with $p_T^j > 20 $ GeV \\[2pt] 
lepton veto $p_T^{e} > 20 $ GeV &$3^\text{rd}$-lepton veto $p_T^{e,\mu} > 10 $ GeV  & number b tags $\geq 2$ \\[2pt] 
lepton veto $p_T^{\mu} > 10 $ GeV &$3^\text{rd}$-lepton veto  $p_T^{\tau} > 18 $ GeV  & $\Delta \phi \left(\text{jet},E_T^{\text{miss}}\right)>$ 1.0 rad \\[2pt]  
jets $\leq 4$ with $p_T^j > 30 $ GeV &$p_T^{\ell\ell} > 60 $ GeV  &  \\[2pt] 
$\Delta \phi \left(\text{jet},p_T^{\text{miss}}\right)>$ 0.4 rad &jets $\leq 1$ with $p_T^j > 30 $ GeV  &  \\[2pt]  
 &top quark veto $p_T^{b}>20$ GeV  & \\[2pt] 
 &$E_T^{\text{miss}} > 100 $ GeV  &   \\[2pt] 
 &$\left|E_T^{\text{miss}} - p_T^{\ell\ell}\right| / p_T^{\ell\ell} < 0.4 $  &   \\[2pt]  
 &$\Delta \phi \left( \ell\ell,\vec{p}_T^{\text{miss}}\right)>$ 2.8 rad  &   \\[2pt]  
 &$\Delta \phi \left(\text{jet},E_T^{\text{miss}}\right)>$ 0.5 rad & \\
\hline 
\end{tabular}
\caption{\label{tab:cuts} Cuts applied in the different search channels, based on the ATLAS mono-jet search \cite{ATLASmono-jets}, and the CMS searches for mono-$Z$ \cite{CMSmono-z} and $t\bar t+E^\text{miss}_T$ final states \cite{Sirunyan:2017xgm}.}
\end{table}

In the following, we discuss the reach of LHC searches for Dark Matter in searches for mono-jet, mono-$Z$ and $t \bar t + E_T^\text{miss}$ final states. We define two benchmark sets of parameters based on the results of the previous sections,
\begin{align}\label{eq:bench1}
&\hspace{-.2cm}\text{Benchmark 1}\hspace{.3cm} M_A\!=\!160\,\text{GeV}, \, M_H\!=\!M_{H^\pm}\!=\!500\,\text{GeV},\, m_\chi\!=\!70\,\text{GeV}, C_5=0.37\,,C_\chi=0\,,\\
&\hspace{-.2cm}\text{Benchmark 2}\hspace{.3cm} M_A\!=\!250\,\text{GeV}, \, M_H\!=\!M_{H^\pm}\!=\!500\,\text{GeV},\,m_\chi\!=\!100\,\text{GeV},C_5=0.37,\,C_\chi=0\,.\label{eq:bench2}
\end{align}
The first benchmark allows for $H\to AA$ decays, whereas for the second benchmark, this decay is kinematically forbidden. For the second benchmark the decay $A \to hZ$ is kinematically allowed for $\cos(\beta-\alpha)\neq 0$, which is forbidden for $M_A=160$ GeV. For both sets of parameters, the relic density can be reproduced for a range of values of $\tan\beta$. We want to stress that the results presented in this section are largely independent of the Dark Matter mass $m_\chi$ as long as $m_\chi< M_A/2$.

\subsection{Signal Generation}

Our Monte Carlo simulation is based on an Universal FeynRules Output (UFO) implementation of the simplified model described in Section \ref{eq:2HDMEFT}. We use FeynRules 2 \cite{Alloul:2013bka}, the NLOCT package \cite{Degrande:2014vpa} embedded in FeynArts 3.9 \cite{Hahn:2000kx}, and the implementation of the two Higgs doublet model \cite{Degrande:2014qga}. We export the UFO file to Madgraph5\_aMC@NLO 2.5.5 \cite{Degrande:2011ua, Alwall:2011uj} for calculating the hard matrix elements. For the generation of these processes MADLOOP \cite{Alwall:2014hca} with its OPP integrand reduction method \cite{Mastrolia:2008jb} inherited from CUTTOOLS \cite{Ossola:2007ax} is used. The showering is performed with the Pythia 8.226 \cite{Sjostrand:2007gs} interface for Madgraph, the detector simulation with Delphes 3.4.0 \cite{deFavereau:2013fsa}, and we take the parton distribution function set NNPDF23\_lo\_as\_130 \cite{Ball:2014uwa}. Our results are valid for the narrow width approximation, which is valid throughout the parameter space considered here, for which the total decay width of the pseudoscalar and the heavy scalar are $\Gamma_A\lesssim 1$ GeV and $\Gamma_H\lesssim 90$ GeV for $\cos(\beta-\alpha)=0$ and $\tan\beta \lesssim 2.5$.

\begin{figure}
\begin{center}
\includegraphics[width=.9\textwidth]{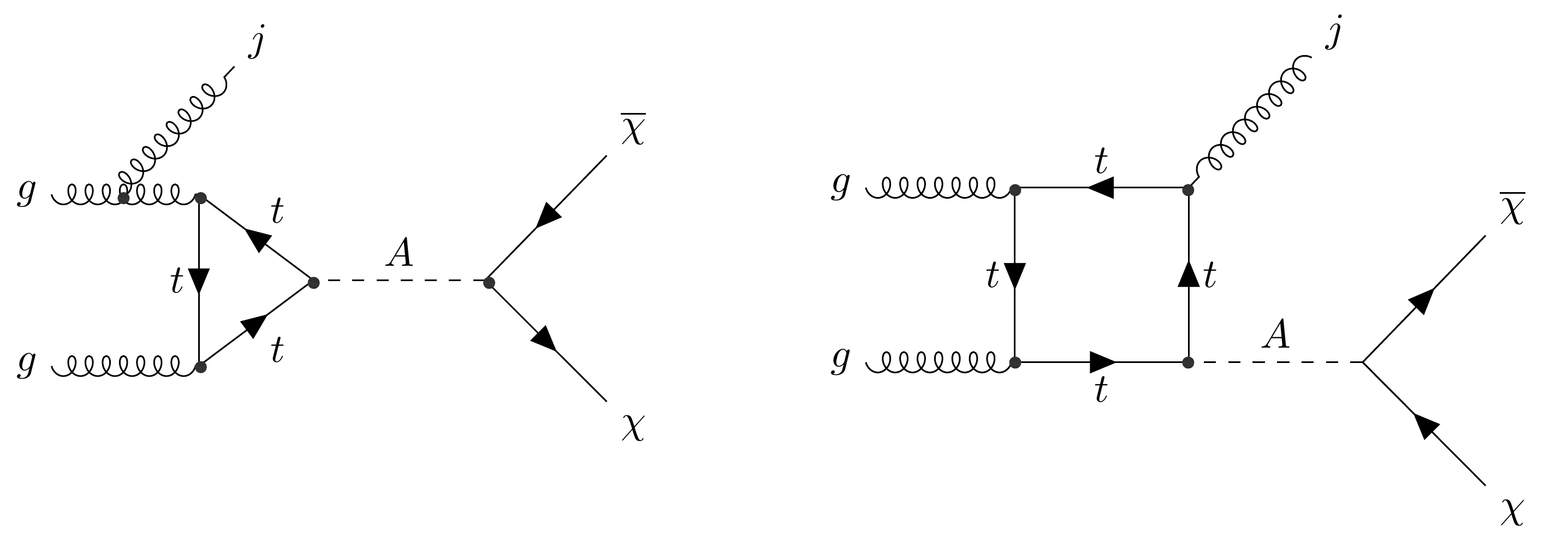} \end{center}\vspace{-.4cm}
\caption{\label{fig:monojet} Diagrams contributing to mono-jet production from initial state radiation. }
\end{figure}

\subsection{Mono-Jets}

The mono-jet signal is generated through initial state radiation and the relevant Feynman diagrams are shown in Figure~\ref{fig:monojet}. Since we concentrate on low values of $\tan\beta$, we neglect the $b$-quark contribution to the gluon fusion loop and generate the signal processes $pp\to A$ and $pp\to A+ j$ with Madgraph5\_aMC@NLO.  We perform a Matrix Element and Parton Shower (ME+PS) merging between the zero and one jet sample, by employing the $k_T$-MLM scheme \cite{Alwall:2007fs, ktmlm} for 0 and 1-jet multiplicities within Pythia8. 
We set the minimal distance in phase space between the QCD partons 
to a quarter of the hard scale in each process. The merging scale 
is chosen to be 1.5 times this distance 
to guarantee a smooth jet measure cutoff. For the rescaling of $\alpha_s$ 5 flavours are taken into account. 
We implement cuts according to the ATLAS mono-jet search \cite{ATLASmono-jets} and validate our results against the rescaled projections of the simplified model used by the LHC Dark Matter Forum (DMF) \cite{ Abercrombie:2015wmb}. The applied cuts are collected in the left column of Table \ref{tab:cuts}. We assume a systematic error of $5\%$ and account for higher order corrections by applying a the mass-dependent N$^2$LO K-factor at $\sqrt{s}=13 $ TeV ranging from 2.1 for $M_A=150$ GeV to 2.37 for $M_A=430$ GeV \cite{Ahmed:2015qda}. 

\begin{figure}
\begin{center}
\includegraphics[width=.9\textwidth]{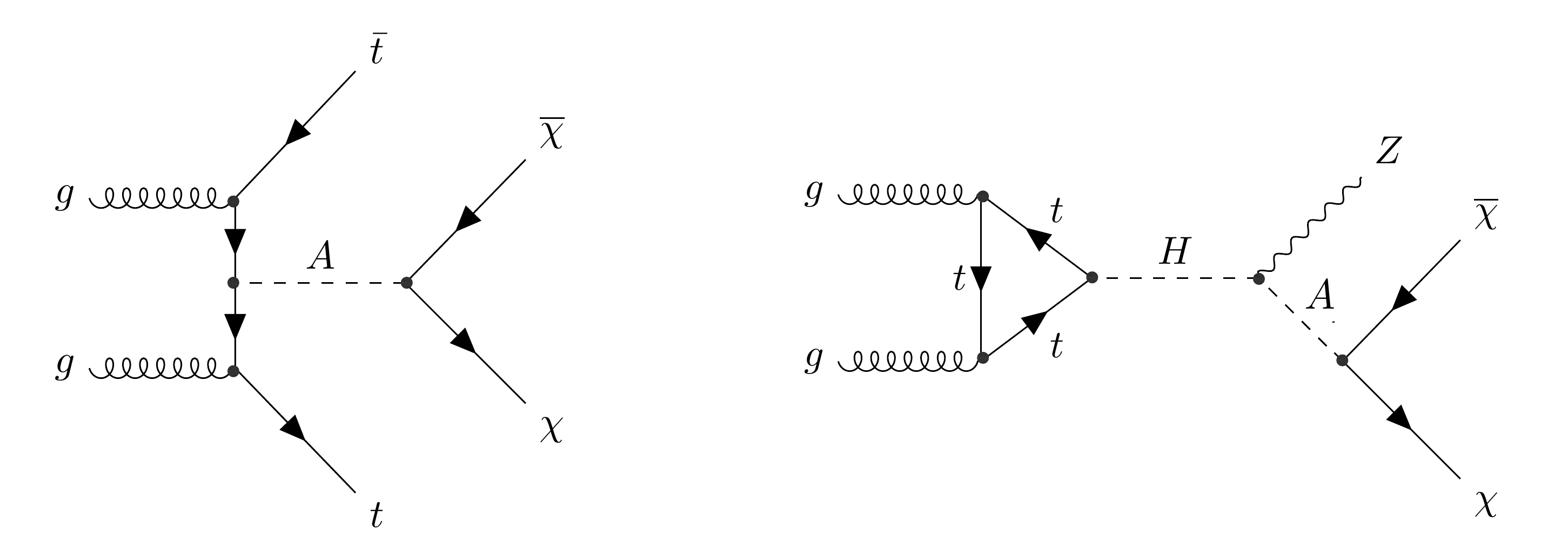} \end{center}\vspace{-.4cm}
\caption{\label{fig:monoZ} Diagrams contributing to $t\bar t+E_T^\text{miss}$ production (left) and resonant mono-$Z$ production (right). }
\end{figure}

\subsection{Mono-Z}
Mono-$Z$ production through initial state radiation is strongly suppressed with respect to the mono-jet and mono-photon final state \cite{Carpenter:2012rg}. The consistency of the pseudoscalar mediator model requires the presence of an additional heavy scalar which cannot be decoupled without violating the stability and unitarity constraints discussed in Section \ref{sec:stab}. This spectrum of heavy scalars in proximity to the pseudoscalar mediator mass allows for a resonantly enhanced mono-$Z$ final state, 
\begin{align}
pp \to H \to AZ \to \chi\bar \chi Z\,,
\end{align}
which we identify as a universal signal of pseudoscalar mediator models. The corresponding Feynman diagram is shown on the right in Figure~\ref{fig:monoZ}. For the parameter space preferred by the constraints in Section \ref{sec:constraints1} and \ref{sec:constraints2}, $\cos(\beta-\alpha)=0$ and $\tan\beta =\mathcal{O}(1)$, we can neglect the contribution from $b$ quarks in the gluon-fusion production of the heavy scalar $H$. The mono-$Z$ signal is therefore directly proportional to the heavy scalar production cross section. We generate the signal at LO with Madgraph5\_aMC@NLO and consider leptonic decays into the $Z$ boson. We apply the cuts used in the CMS mono-$Z$ search \cite{CMSmono-z} collected in the center coloumn of Table~\ref{tab:cuts}. The implementation of the cuts is validated against the dominant irreducible background process $pp\to ZZ \to \nu\bar\nu \ell^+\ell^-$. In producing the exclusion limits, we assume a systematic error of $10\%$ and account for higher order corrections by applying an N$^2$LO K-factor of 2.3 for $M_H=500$ GeV at $\sqrt{s}=13$ TeV \cite{Ahmed:2015qda, Harlander:2002wh}.

\begin{figure}
\begin{center}
\includegraphics[width=1\textwidth]{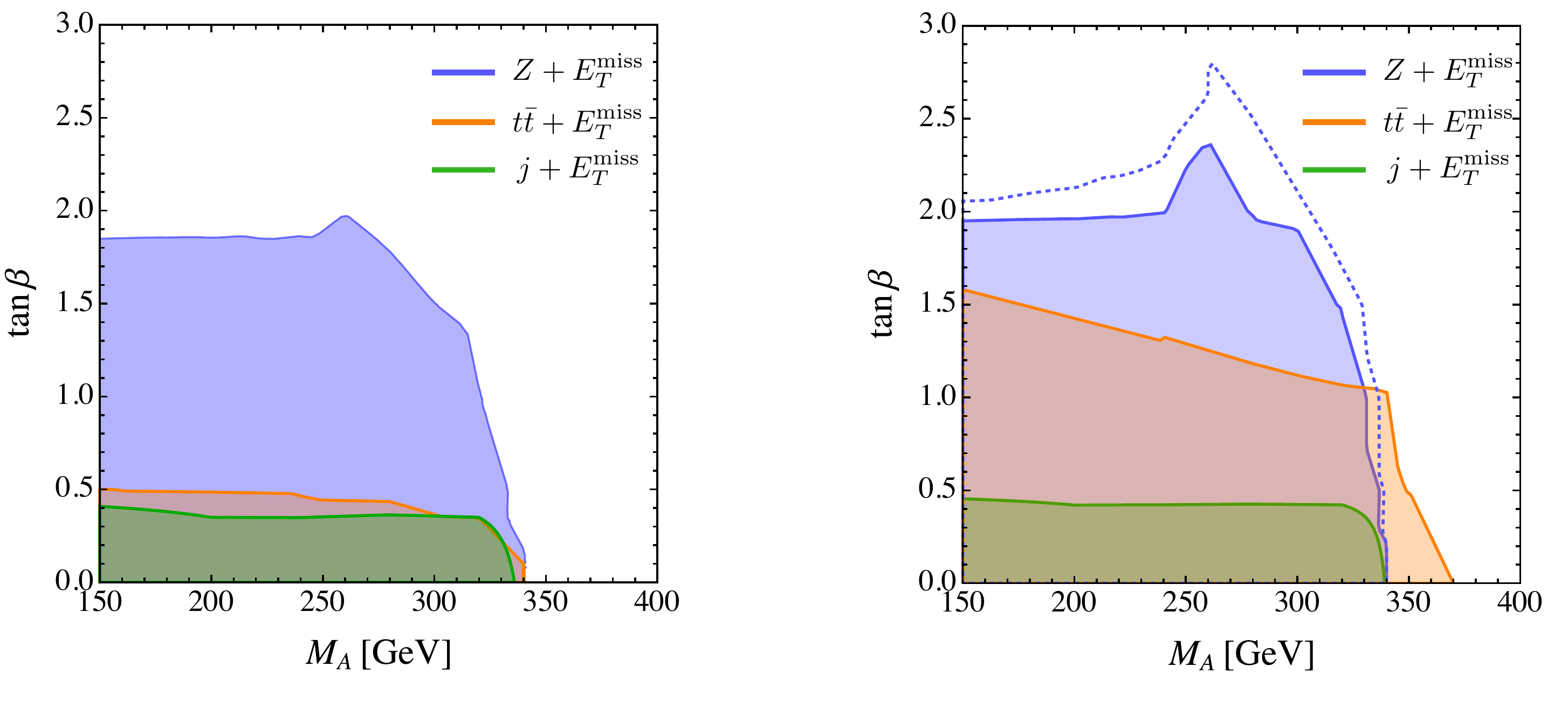} \end{center}\vspace{-.5cm}
\caption{\label{fig:final1}  Left panel: Exclusion contours for different mono-X searches at the LHC. Right panel: Projections for the reach of mono-X searches with $300$ fb$^{-1}$ . }
\end{figure}

\subsection{\boldmath$t\bar t A$ Production }

The flavour-dependent couplings of spin-0 mediators particularly motivate searches for missing energy in associated heavy flavour production. The low values of $\tan\beta\sim \mathcal{O}(1)$ preferred by the class of models discussed here strongly favour the $t\bar t +E_T^\text{miss}$ final state over $b\bar b +E_T^\text{miss}$ production. The corresponding Feynman diagram for the process $pp\to A t\bar t\to t\bar t +E_T^\text{miss}$ is shown on the left of Figure~\ref{fig:monoZ}. We generate the events at LO and apply the cuts given in the right panel of Table~\ref{tab:cuts}. 
We neglect the systematic uncertainty and assume a mass-independent K-factor of $1.1$ for the $t \bar t+A$ production at $\sqrt{s}=13$ TeV \cite{Maltoni:2016yxb}. \\

\subsection{Other Mono-X Signatures}
Besides mono-jet, mono-$Z$ and $t\bar t+E_T^\text{miss}$ searches, colliders can search for Dark Matter in mono-photon, mono-$W$ and mono-Higgs final states. In the model discussed here, the production of a Higgs or photon in association with the pseudoscalar $A$ occurs through initial state radiation. Mono-photon production is therefore suppressed by $Q_f^2N_C\alpha_e/\alpha_s$ with respect to the mono-jet signal, whereas Higgs radiation from the top loop requires the production of two massive spin-0 bosons. The corresponding cross sections are negligibly small compared to the mono-jet, mono-$Z$ and $t\bar t+A$ production. We emphasize that this hierarchy of signatures can be different in UV completions resolving the effective coupling of the pseudoscalar to Dark Matter. The extension by an additional light pseudoscalar singlet for example results in a striking resonant mono-Higgs signal \cite{Bauer:2017ota}. In contrast, a mono-$W$ final state can be resonantly produced through $pp\to H^+\to W^+ A\to W^+ \bar \chi \chi$. From Figure~\ref{fig:HpmBRs} follows, that the branching ratio $\text{Br}(H^\pm \to W^\pm A)$ is large in the interesting window of $1\lesssim \tan\beta \lesssim 10$. However, the production rate of the charged scalar, $\sigma(pp \rightarrow H^{\pm})\approx 0.5 $ fb and $\sigma(gg \rightarrow W^- H^+)\approx 0.01 $ pb is considerably smaller than $\sigma(gg \rightarrow H)\approx 1.77 $ pb, leading to a much smaller cross secion $\sigma(p p \to H^+ \to W^+ +E_T^\text{miss})$. A more relevant signal arises from charged Higgs production in association with heavy flavour, $\sigma(g b \rightarrow H^{-}t)\approx 0.17$ pb. Recently, an analysis of single-top production in association with missing energy $pp\to t W^\pm + E_T^\text{miss}$ has been performed for pseudoscalar mediators in two Higgs doublet models \cite{Pani:2017qyd}. Depending on the mass hierarchy this search can be competitive with the mono-$Z$ channel, in particular if $H \to AA$ is the dominant decay channel of the heavy neutral Higgs.

\begin{figure}
\begin{center}
\includegraphics[width=1.\textwidth]{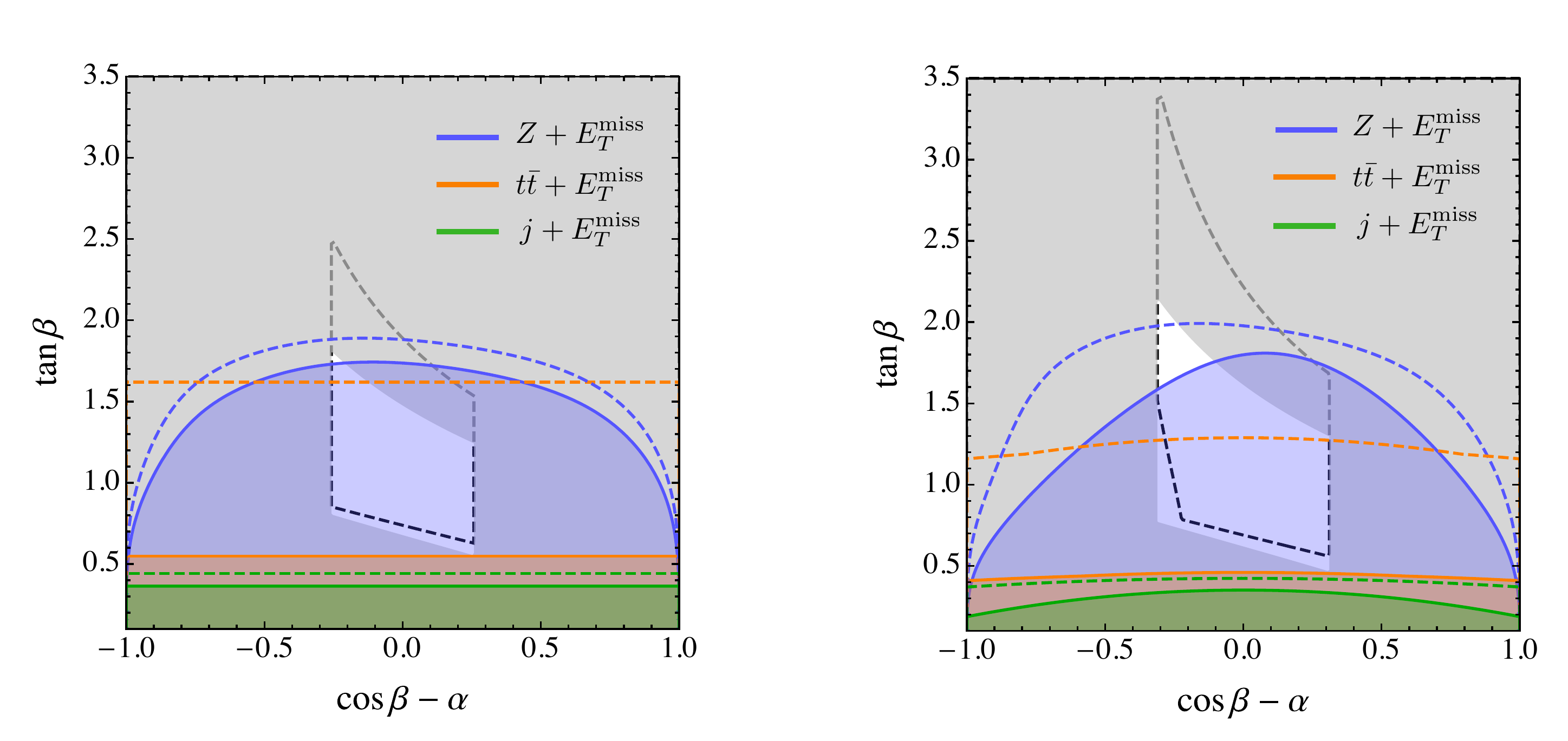} \end{center}\vspace{-.5cm}
\caption{\label{fig:final2}
Reach of the search for mono-$Z$ (blue), mono-jet (green) and $t\bar t+E_T^\text{miss}$ (orange) for the benchmarks defined in \eqref{eq:bench1}  (left panel) and \eqref{eq:bench2} (right panel). Dashed contours correspond to the projected reach for a luminosity of $300$ fb$^{-1}$ and the gray shaded region outside the white shaded area (outside the black dashed contours) is excluded by constraints from unitarity, the stability of the potential and electroweak precision constraints for a potential with $\lambda_5=\lambda_6=\lambda_7=0$ ($\lambda_5=\lambda_7=0, \lambda_6=1$).}
\end{figure}

\subsection{Discussion}
In Figure~\ref{fig:final1}, we display the reach of current and future searches for Dark Matter produced at the LHC mediated by a pseudoscalar in the mono-$Z$ (blue), mono-jet (green) and $t\bar t + E_T^\text{miss}$ (orange) final state in the $M_A-\tan\beta$ plane. We have fixed $M_H=M_{H^+}=500\, \text{GeV}, \cos(\beta-\alpha)=0$ and $c_5=1.5\times 10^{-3} (\Lambda/\text{GeV}), c_\chi=0$. The reach of the mono-jet and $t\bar t+E_T^\text{miss}$ searches are limited, because it is suppressed by phasespace and the $t \bar t A$ coupling that scales like $1/\tan^4\beta$ and loose sensitivity above $\tan \beta = 0.4$ and $0.5$, respectively. The resonant production $p  p \to H \to A Z$ is sensitive up to values of $\tan \beta \approx 1.85$ for $M_A=150$ GeV. All three search channels loose sensitivity above $M_A= 2m_t$, where the branching ratios of both the scalar $H$ and the pseudoscalar $A$ into $t \bar t-$pairs dominate. The exclusion contours from mono-jet and $t \bar t+ E_T^\text{miss}$ searches fall more steeply compared to the mono-$Z$ exclusion region, because the Br$(H \to t \bar t)$ is already relevant for masses $M_A< 2 m_t$ and for $M_A\gtrsim M_H/2$, and the sensitivity grows at $M_H\approx 2 M_A$, because the $H\to AA$ decay is kinematically forbidden. In the right panel of Figure~\ref{fig:final1}, the future reach of the different searches is shown for a luminosity of $300$ fb$^{-1}$. The $t\bar t+E_T^\text{miss}$ search is currently statistically limited and is therefore expected to improve the most with increased statistics, whereas improvements in the mono-$Z$ and mono-jet channels are conditional to reducing the systematic uncertainties.
We illustrate this by also giving the future reach for $300$ fb$^{-1}$ and a systematic error of $6\%$ for the mono-$Z$ final state (given by the blue, dotted contour).\\
In Figure~\ref{fig:final2}, we display the reach of the three different mono-X searches for the two benchmark parameter sets defined in \eqref{eq:bench1} and \eqref{eq:bench2} in the $\cos(\beta-\alpha)-\tan\beta$ plane. Dashed, colored contours correspond to the projected reach for a luminosity of $300$ fb$^{-1}$. The gray shaded area is disfavoured by stability and electroweak precision constraints, leaving a window around $\cos(\beta-\alpha)=0$ and $\tan\beta=1$. We have emphasized throughout the discussion that these constraints should not be taken at face value, because they are subject to change if the simplified model is UV completed. It is remarkable however that the mono-$Z$ search can cover a large range of the parameter space motivated by the constraints in the simplified model. The parameter space allowed by indirect constraints can increase if quartic Higgs couplings beyond $\lambda_1-\lambda_4$ are allowed and the parameter space within the black dashed contours corresponds to the allowed region for a value of $\lambda_6=1$. For $|\cos(\beta-\alpha)|> 0$ and $M_A=160$ GeV (left panel), the reach of the mono-$Z$ search drops, because of the parametric dependence of the width $\Gamma(H \to AZ)\propto \sin^2(\alpha-\beta)$. In contrast, the reach of the mono-jet and $t\bar t+E_T^\text{miss}$ search is constant with $\cos(\beta-\alpha)$. For $M_A=250$ GeV (right panel), the mono-$Z$ reach drops more rapidly for sizable $\cos(\beta-\alpha)$ and the reach of the mono-jet and $t\bar t+E_T^\text{miss}$ search drop as well. The reason is the $A\to Z h$ decay channel, which opens up for $M_A\gtrsim M_h+M_Z$ and scales like $\Gamma(A\to hZ)\propto \cos^2(\beta-\alpha)$. Future searches for the mono-$Z$ final state can rule out almost the complete parameter space of the simplified model and provide the best channel to search for more complete models in the absence of additional light mediator states.

\section{Conclusions}\label{sec:conclusion}
The LHC is particularly powerful in probing pseudoscalar mediators to a dark sector, which are notoriously challenging for direct detection experiments and provide interesting signatures for searches for indirect signs of Dark Matter \cite{Izaguirre:2014vva}. In contrast to scalar or vector mediators, pseudoscalars can not couple to the SM through renormalizable mixing terms including either the SM Higgs or the hypercharge gauge boson. In order to be consistent, models of pseudoscalar mediators therefore require additional fields beyond the mediator and a dark matter candidate. There is no unique way to economically fix this additional particle content, leading to  different models whose phenomenology strongly depend on this choice. We present an analysis of universal  signals of these various models, obtained by allowing for renormalizable couplings to SM fields through embedding the pseudoscalar in a two Higgs doublet model, but with effective operator couplings to the Dark Matter candidate $\chi$. This effective coupling can be understood as the limit in which additional fermions or scalars beyond the mediator multiplet are integrated out. There is therefore a straightforward way to match UV completions with a more complicated mediator sector \cite{Goncalves:2016iyg, Bauer:2017ota, Baek:2017vzd} as well as UV completions with the Dark Matter candidate as a component of an electroweak multiplet \cite{Berlin:2015wwa} to the model discussed here. \\
We find that a resonant mono-$Z$ signal through $pp\to H \to AZ \to \bar \chi \chi Z$ is a striking, universal signal of all these pseudoscalar mediator models in the parameter space in which mono-X searches are most powerful. \\
Direct Detection experiments and measurements of Higgs couplings at the LHC provide strong limits on the couplings of the scalars $h$ and $H$ to Dark Matter, motivating a purely pseudoscalar coupling to Dark Matter. The pseudoscalar $A$ therefore is the only mediator between the SM and the dark sector, unless the scalar potential is CP-violating. 
The dominant branching ratio of the heavy scalar Higgs is then $\text{Br}(H \to A Z)$, if it is kinematically allowed. The pseudoscalar dominantly decays into Dark Matter as long as the decay channel of $A$ into Dark Matter only competes with the $A\to b\bar b$ decay mode, that is for $M_A< 2 m_t$. For pseudoscalar masses close to the mass of the scalar, the $H \to A Z$ decay channel is kinematically forbidden, but indirect constraints from measurements of flavour changing neutral currents and electroweak precision observables impose a general limit of $M_{H}\approx M_{H^\pm}\gtrsim 500$ GeV (flavour constraints are weaker in the case of a 2HDM of type I). If both these conditions are met, $A\to t\bar t$ becomes the dominant decay channel of the pseudoscalar and the reach of mono-X searches is severely diminished. The decay mode $A \to h Z$ is accessible if $M_A\gtrsim M_h+M_Z$ and $\cos(\beta-\alpha)\neq 0$. Values of $|\cos(\beta-\alpha)|>0$ are however strongly constrained by Higgs coupling strength measurements, resulting in large pseudoscalar decay widths into Dark Matter even if this channel is kinematically allowed. In contrast to the resonant mono-$Z$ final state in this model, mono-jet production occurs through initial-state radiation.
As a result, we find that both mono-jet and associated production of the mediator with a $t\bar t$ pair are not resonantly enhanced and the reach of LHC searches in these channels is more limited. \\
Interestingly, the parameter space preferred by electroweak precision observables, a stable minimum of the scalar potential and the unitarity of scalar scattering amplitudes overlaps with the parameter space for which the correct relic density of Dark Matter can be reproduced in this model. The combination of these indirect constraints therefore leaves a well motivated window of parameter space. Future mono-$Z$ searches at the LHC will be able to probe almost this entire window.\\
Additional degrees of freedom that are expected in UV completions of this model will extend this parameter space and could provide additional signatures such as mono-Higgs final states or associated production of the Dark matter candidate with its charged partners, but the mono-$Z$ final state remains a universal signal of consistent pseudoscalar mediator models unless new light particles are present in the model. \\
During final preparations of this paper, an analysis of the $pp \to t W^- + E_T^\text{miss} $ final state appeared that provides constraints on the $H^\pm\to t W^\pm A\to t W^\pm \chi \bar \chi$ decay in this model could be competitive with the reach in the mono-$Z$ channel \cite{Pani:2017qyd}.

\section*{Acknowledgements}
We acknowledge helpful discussions with members of the DFG Research Unit 2239: "New Physics at the LHC" during meetings and workshops. Martin Bauer thanks Uli Haisch for helpful comments. During the final preparation  Valentin Tenorth acknowledges financial support by the IMPRS-PTFS. 

\begin{appendix}

\section*{Appendix : Decay Widths}\label{sec:pw}
In the following, we collect the partial decay widths for the heavy scalar $H$, the charged scalar $H^\pm$ and the pseudoscalar $A$ in the alignment limit $\cos (\beta-\alpha) =0$. Decays that only become relevant for $\cos(\beta-\alpha)\neq0$ have not been shown, but can be found for example in \cite{Gunion:1989we}. For the scalar $H$, one has
\begin{align}
\Gamma(H \to t\bar t)&=\frac{3}{8\pi}\frac{m_t^2}{v^2}\kappa_{H_u}^2 M_H \bigg(1-\frac{4m_t^2}{M_H^2}\bigg)^{3/2}\,,\\
\Gamma(H \to \chi\bar \chi)&=\frac{1}{8\pi}g_{H\chi}^2 M_H \left(1-\frac{4m_\chi^2}{M_H^2}\right)^{3/2}\,,\\
\Gamma(H \to ZA)&=\frac{1}{16\pi}s_{\beta-\alpha}^2\frac{M_Z^4}{v^2M_H}\,\lambda(M_A^2,M_Z^2,M_H^2)^{1/2}\,\lambda(M_A^2,M_H^2,M_W^2)\,,\\
\Gamma(H \to AA)&=\frac{1}{32\pi}\frac{1}{v^2M_H}\bigg(M_H^2\,c_{\beta-\alpha}+(M_H^2-M_A^2)s_{\beta-\alpha}\Big(t_\beta-\frac{1}{t_\beta}\Big)\bigg)^2\left(1-\frac{4M_A^2}{M_H^2}\right)^{1/2}\,,
\end{align}
where $\lambda(x,y,z)=((x+y-z)^2-4xy)/z^2$ and the couplings $\kappa_{H_u}$ and $g_{H\chi}$ are defined in Table~\ref{tab:kappas} and \eqref{eq:DMcouplings}. For the pseudoscalar $A$, the following partial decay widths are relevant
\begin{align}
\Gamma(A \to t\bar t)&=\frac{3}{8\pi}\frac{m_t^2}{v^2}\kappa_{A_u}^2 M_A \left(1-\frac{4m_t^2}{M_A^2}\right)^{1/2}\,,\\
\Gamma(A \to b\bar b)&=\frac{3}{8\pi}\frac{m_b^2}{v^2}\kappa_{A_d}^2 M_A \left(1-\frac{4m_b^2}{M_A^2}\right)^{1/2}\,,\\
\Gamma(A \to \tau^+ \tau^-)&=\frac{1}{8\pi}\frac{m_\tau^2}{v^2}\kappa_{A_\ell}^2 M_A \left(1-\frac{4m_\tau^2}{M_A^2}\right)^{1/2}\,,\\
\Gamma(A \to \chi\bar \chi)&=\frac{1}{8\pi}g_{A5}^2 M_A \left(1-\frac{4m_\chi^2}{M_A^2}\right)^{1/2}\,,
\end{align}
and $\Gamma(A\to c \bar c)$ follows from $\Gamma(A\to t \bar t)$ with the replacement $m_t\to m_c$. For the charged scalar $H^\pm$, one has
\begin{align}
\Gamma(H^+ \to t\bar b )&=\frac{3}{8\pi}\frac{|V_{tb}|^2}{M_{H^\pm}v^2}\lambda(m_t^2,m_b^2,M_{H^\pm}^2)^{1/2}  \!\left((M_{H^\pm}^2\!-\!m_t^2\!-\!m_b^2)(m_b^2 \kappa_{A_d}^2\!+\!m_t^2\kappa_{A_t}^2)\!-\!4m_t^2m_b^2\right)\,,\\
\Gamma(H^+ \to \tau^+ \nu )&=\frac{1}{8\pi}\frac{1}{M_{H^\pm}v^2}m_\tau^2\kappa_{A_\ell}^2\left(1-\frac{m_\tau^2}{M_{H^\pm}^2}\right)^3\,,\\
\Gamma(H^+ \to AW^+)&=\frac{1}{16\pi c_W^2}\frac{M_W^4}{M_{H^\pm}\,v^2}\lambda(M_A^2,M_W^2,M_{H^\pm}^2)^{1/2}\lambda(M_A^2,M_{H^\pm}^2,M_W^2)\,.
\end{align}

\end{appendix}

\end{document}